\documentclass[a4paper,12pt]{article}
\usepackage[english]{babel}
\usepackage{graphicx,color}
\usepackage{amsmath,amsthm,amssymb,amsfonts}
\usepackage{amsthm}
\usepackage{mathtext}
\usepackage{exscale}
\usepackage{latexsym}
\usepackage[all]{xy}
\usepackage{epsfig}
\usepackage{lmodern}
\usepackage[T1]{fontenc}
\usepackage{textcomp}
\usepackage[cp1250]{inputenc}
\usepackage[bookmarks,colorlinks]{hyperref}

\hypersetup{colorlinks,%
linkcolor=red,%
urlcolor=blue,%
citecolor=red,%
filecolor=magenta}

\title{Quantum gravity as the way from spacetime to space quantum states thermodynamics}

\author{\textbf{Lukasz A. Glinka}$^{1,2}$\footnote{E-mail to: ~\href{mailto:laglinka@gmail.com}{\texttt{laglinka@gmail.com}}~,~\href{mailto:glinka@theor.jinr.ru}{\texttt{glinka@theor.jinr.ru}}}\vspace*{10pt}\\
$^{1}$\normalsize{\emph{Bogoliubov Laboratory of Theoretical Physics}},\\
\normalsize{\emph{Joint Institute for Nuclear Research}},\\
\normalsize{\emph{Joliot--Curie 6, 141980 Dubna, Moscow Region, Russia}}\vspace*{10pt}\\
$^{2}$\normalsize{\emph{Universit$\grave{\mathit{a}}$ degli Studi di Udine,}}\\
\normalsize{\emph{Dipartimento di Matematica e Informatica,}}\\
\normalsize{\emph{via delle Scienze, 206 33100 UDINE (UD) -Italy}}}
\date{\today}

\newtheorem*{conjecture}{Conjecture}
\newcommand{\tr}{\mathrm{Tr}}

\begin{document}

\maketitle
\thispagestyle{empty}
\begin{abstract}

Physical spacetime geometry follows from some effective thermodynamics of quantum states of all fields and particles described in frames of General Relativity. In the sense of pure field theoretical Einstein's point of view on gravitation the thermodynamic information is actually quantum gravity. We propose new realization of this old idea by studying the canonical $3+1$ Dirac--ADM approach to pseudo--Riemannian (Lorentzian) manifold of General Relativity. We derive the Wheeler--DeWitt theory as the Global One-Dimensional classical field theory of the Bose field associated with embedded 3-space, where Wheeler's superspace metric is absent. The classical theory is discussed, some deductions on tachyon state, Dark Energy density and cosmological constant are included. Reduction to 1st order evolution is carried out, and quantum theory by the second quantization in the Fock space of creators and annihilators is constructed by employing the Heisenberg equation and the bosonic Bogoliubov transformation for diagonalization. In result we find the static reper with stable vacuum, where quantum states of 3-space can be considered, and finally space quantum states thermodynamics is formulated.
\end{abstract}
\newpage
\tableofcontents

\newpage

\section{Einstein's Thermodynamic Legacy}

\begin{flushleft}
\textit{Thermodynamics is the only physical theory of universal content concerning which I am convinced that, within the framework of the applicability of its basic concepts, it will never be overthrown.}
\end{flushleft}
\noindent These words of Dr. Albert Einstein written in his autobiographical notes \cite{ein1} are the testament of his life in Science. One hundred years after Einstein's discoveries, in the centaury of microcomputers, the testament sounds strange. Today theoretical as well as mathematical physics treats thermodynamical investigations with some very subtle kind of contempt. In common established conviction thermodynamics is clear and well understood branch of physics, and is need only for engineering sciences, but not for theoreticians and mathematicians -- this branch is categorized by more technologic than scientific debatable level.

However, from methodological point of view, it is interesting that the General Relativity founder was the well defined thermodynamic physicist, and in spite of this crucial fact he successfully formulated physical as well as mathematical fundamentals on the modern view on gravitation, so that General Relativity today has a status of a physical theory well confirmed by experimental data from the nearest regions of Cosmos.

As it is commonly known, towards the end of his life Einstein did general field theoretical considerations about physics. The reason of the general field theory dream seems to be very simple as well as very complex. Namely, from the historical point of view, the man who gave contributions to theory of diffusion processes and explained unexpected difference between theoretical and experimental predictions in the photoelectrical effect of falling down light quanta on a metal, surprisingly gave simultaneously and practically in the same time the crucial investigation to contemporary thinking about gravitational phenomena on classical level as, \emph{i.e.} he computed theoretical values of precession of the perihelion of Mercury and binding of light rays around the sun, and generalizes the Newton law of universal gravitation. These regions of scientific activity at first glance lie in the most far conceptual points, and are not connected by straight intellectual line. Can it be only a coincidence that these outermost points was present in the Einstein considerations? Maybe it is not obvious, but really it seems that Dr. Einstein discovered only the one universal true, namely that \emph{thermodynamics is an essence of all physics}, because really thermodynamic effects are present on \emph{experimental level}.

In this manner also gravitation, by its obvious presence in Nature and deep physical consequences can be also considered as an thermodynamic effect. Really, the main goal of the Einstein vision was understanding that \emph{it should be possible to describe geometry by thermodynamics}. By this reason, thermodynamics realizes the concept of the \emph{general field theory} searched by Einstein.

In this paper we will derive some new realization of the Einstein idea. This is a mathematical way between a pseudo-Riemannian manifold given by spacetime of General Relativity, and thermodynamics of quantum states of this spacetime. Existence of this way is not obvious, and by this we propose to take into considerations the following point of view
\begin{conjecture}
Gravitation determined by General Relativistic spacetime can be described as the effective quantum states thermodynamics.
\end{conjecture}
\noindent The realization of this conjecture is essentially contained in the idea called in this paper \emph{Thermodynamical Einstein's Dream}. This idea is comprehend as some hypothetical way from the classical object that is a pseudo-Riemannian manifold defined by four dimensional spacetime metrics being a solution of the Einstein--Hilbert field equations of General Relativity, to generalized thermodynamics of the Bose gas of quantum states of 3-dimensional space. The conception of space quantum states arises naturally from the canonical $3+1$ Dirac--ADM approach to General Relativity, that determines theory of gravitation as an time evolution of 3-dimensional geometry of embedded space. Conceptually the thermodynamics realizes the quantum theory of general gravitational fields in the strict sense of general field theory. In this manner, this paper is devoted to give the new proposal for quantum gravity realization.

Contents of this paper is as follows. First, we recall very briefly basic classical $3+1$ approach to the Einstein--Hilbert General Relativity \cite{ein,hil} called geometrodynamics, that is studying of the splitting of 4-dimensional pseudo-Riemannian geometry into 1-dimensional time and 3-dimensional space, and interprets of General Relativity as time evolution of the embedded 3-space geometry. This canonical $3+1$ Arnowitt--Deser--Misner approach to General Relativity \cite{adm,per} leads immediately to the Hamiltonian secondary constraint of General Relativity given by some generalized the Einstein--Hamilton--Jacobi equation, that after application of the Dirac first quantization method \cite{dir} gives in the result the Wheeler--DeWitt equation and leads to the conception of the Wheeler's superspace \cite{whe,dew,haw} as the configurational space of General Relativity. The superspatial quantum equation of motion mathematically is the second order functional differential equation with respect to metrics of 3-dimensional space, and is commonly interpreted \cite{haw,abl,rov} as a kind of the nonrelativistic Schr\"odinger quantum mechanics in superspace for the Wheeler--DeWitt wave function that is a functional of embedded 3-space metric field.

However, by the relativistic character of General Relativity, in this paper it is proposed to reinterpret this equation not as nonrelativistic equation but as relativistic one, that is the Klein--Gordon--Fock equation for some abstract Bose field. It is demonstrated explicitly that in frames of $3+1$ metric field decomposition one can get rid of the superspace metric from the Wheeler--DeWitt equation and to change differentiation from metrics of 3-dimensional space onto determinant of 3-metrics, and in result treat the Wheeler--DeWitt wave function as the one dimensional Bose field associated with a pseudo--Riemannian manifold of General Relativity by three dimensional geometry of embedded space. Some aspects of classical theory of the boson are developed and discussed in this paper, deductions for the tachyon state, Dark Energy and cosmological constant value are included. The classical field theory is quantized by employing the second quantization in form of generalized canonical commutation relations agreed with general Von Neumann--Araki--Woods algebraic approach \cite{neu,aw} and the correct Fock space is builded. By nonlinear character of equations of motion in the Fock space, diagonalization procedure based on the Heisenberg operator evolution with using of the bosonic Bogoliubov transformation \cite{bos} is proposed, and in the result the fundamental functional operator Fock reper associated to initial data is obtained, where stable quantum vacuum state is naturally present. Quantum states of 3-dimensional space, called here Space Quantum States (SQS), are defined with respect to this static reper, and thermodynamics of the Bose gas of space quantum states is formulated by application of the one-particle density matrix method in this static basis. The SQS system is analyzed from point of view of one-point correlations of the Bose field, and thermodynamically stable phase of the system in the limit of huge number of quantum states produced from stable Bogoliubov vacuum is chosen as the correct in thermodynamical equilibrium sense. Fundamental thermodynamic characteristics are computed. The equipartition law is used to obtain a number of degrees of freedom in the classical limit of the Bose gas, and spacetime coordinates are interpreted as four thermodynamical degrees of freedom. Entropy of the Bose gas of space quantum states is analyzed, and roles of initial data vacuum quantum states, and the Cold Big Bang of SQS from vacuum are discussed. In this manner the quantum theory of gravitation is realized as generalized thermodynamics of the Bose gas of space quantum states.
\newpage
\section{Hamiltonian $3+1$ Quantum Gravity}
In this section we present some standard results which have a basic status for General Relativity and quantum gravity, and are need for further developments of this paper. This is the Arnowitt--Deser--Misner canonical $3+1$ approach to a Lorentzian manifold given by a solution of the Einstein--Hilbert field equations of General Relativity and its Dirac's primary quantization that leads to the Wheeler--DeWitt equation and the concept of Superspace.
\subsection{The Einstein--Hilbert field equations}
General Relativity can be obtained from the four-geometry action with fixed on a boundary three-geometry \cite{haw,mtw} \footnote{In this paper the units $8\pi G/3=c=\hbar=k_B=1$ are used.}
\begin{equation}\label{eh0}
S[g]=\int_{M}d^4x\sqrt{-g}\left\{-\dfrac{1}{6}R[g]+\dfrac{\Lambda}{3}+\mathcal{L}\right\}-\dfrac{1}{3}\int_{\partial M}d^3x\sqrt{h}K[h],
\end{equation}
where $(M,g)$ is a pseudo--Riemannian manifold \cite{rie} with a boundary $(\partial M,h)$, $h=\det{h_{ij}}$ is 3-volume form, $K[h]=\tr K_{ij}$ is traced the second fundamental form, related to unit normal vector $n^i$ by $K_{ij}=-\nabla_{(i}n_{j)}$ and called the extrinsic Gauss--Codazzi curvature (see \emph{e.g.} \cite{pet}) of a boundary, $g=\det{g_{\mu\nu}}$ is 4-volume form, $R[g]$ is the Ricci scalar curvature, $\Lambda$ is cosmological constant, and $\mathcal{L}$ is a lagrangian of all physical fields considered on a manifold, called Matter Lagrangian. By application of the Hilbert--Palatini variational principle \cite{hil,pal} with respect to the fundamental field $g_{\mu\nu}$ to the action (\ref{eh0})
\begin{equation}\label{eh1}
\dfrac{\delta S[g]}{\delta g_{\mu\nu}}=0,
\end{equation}
with boundary condition
\begin{equation}
  \delta S[g]\left|_{\partial M}\right.=0,
\end{equation}
one can obtain the Einstein--Hilbert field equations of General Relativity
\begin{equation}\label{gr}
R_{\mu\nu}-\dfrac{R}{2}g_{\mu\nu}+\Lambda g_{\mu\nu}=3T_{\mu\nu},
\end{equation}
where $T_{\mu\nu}$ is the stress--energy tensor
\begin{equation}\label{eh3}
T_{\mu\nu}=\frac{2}{\sqrt{-g}}\frac{\delta\left(\sqrt{-g}\mathcal{L}\right)}{\delta g^{\mu\nu}},
\end{equation}
$R_{\mu\nu}$ is the Ricci curvature tensor that is contracted the Riemann--Christoffel curvature tensor $R^\lambda_{\mu\alpha\nu}$, and is dependent on the Christoffel affine connections $\Gamma^\rho_{\mu\nu}$ and their coordinate derivatives
\begin{eqnarray}\label{eh2}
R^\lambda_{\mu\alpha\nu}&=&\Gamma^\lambda_{\mu\nu,\alpha}-\Gamma^\lambda_{\mu\alpha,\nu}+\Gamma^\lambda_{\sigma\alpha}\Gamma^\sigma_{\mu\nu}-\Gamma^\lambda_{\sigma\nu}\Gamma^\sigma_{\mu\alpha},\\
R_{\mu\nu}&=&R^\lambda_{\mu\lambda\nu}=\Gamma^\lambda_{\mu\lambda\alpha}-\Gamma^\lambda_{\mu\alpha,\lambda}+\Gamma^\lambda_{\sigma\alpha}\Gamma^\sigma_{\mu\lambda}-\Gamma^\lambda_{\sigma\lambda}\Gamma^\sigma_{\mu\alpha},~~R=g^{\mu\nu}R_{\mu\nu}\\
\Gamma^\rho_{\mu\nu}&=&\dfrac{1}{2}g^{\rho\sigma}\left(g_{\mu\sigma,\nu}+g_{\sigma\nu,\mu}-g_{\mu\nu,\sigma}\right),\label{chris}
\end{eqnarray}
where holonomic basis \cite{mtw} was chosen.

\subsection{$3+1$ General Relativity}
Let us introduce coordinate system chosen by the condition so that boundary space is a constant time $t$ surface and write the spacetime metric field being a solution of the Einstein--Hilbert field equations (\ref{gr}) in the following way
\begin{eqnarray}\label{adm}
ds^2&=&g_{\mu\nu}dx^\mu dx^\nu=-N^2dt^2+h_{ij}\left(dx^i+N^idt\right)\left(dx^j+N^jdt\right)=\nonumber\\
&=&-\left(N^2-N_iN^i\right)dt^2+N_idx^idt+N_jdx^jdt+h_{ij}dx^idx^j,
\end{eqnarray}
that actually is the Pythagoras' theorem between two points lying on two distinguish constant time (spacelike) hypersurfaces, and was firstly investigated by Arnowitt, Deser and Misner (ADM) \cite{adm}. By this four-dimensional metrics $g_{\mu\nu}$ of the Einstein--Hilbert General Relativity Riemannian manifold in the canonical $3+1$ ADM approach has the following form
\begin{eqnarray}\label{dec}
  &&g_{\mu\nu}=\left[\begin{array}{cc}-N^2+N_iN^i&N_j\\N_i&h_{ij}\end{array}\right],\\
  &&g^{\mu\nu}=\left[\begin{array}{cc}-\dfrac{1}{N^2}&\dfrac{N^j}{N^2}\vspace*{5pt}\\ \dfrac{N^i}{N^2}&h^{ij}-\dfrac{N^iN^j}{N^2}\end{array}\right],\\
  &&h_{ik}h^{kj}=\delta_i^j,~~N^i=h^{ij}N_j,~~g=N^2h,
\end{eqnarray}
In this case the action (\ref{eh0}) becomes
\begin{eqnarray}\label{gd}
  S[g]=\int dt\left[\int_{\partial M} d^3x\left\{\pi\dot{N}+\pi^i\dot{N_i}+\pi^{ij}\dot{h}_{ij}-NH-N_iH^i\right\}\right],
\end{eqnarray}
where
\begin{eqnarray}
H&=&\sqrt{h}\left\{(K^i_i[h])^2-(K^2[h])^i_i+R[h]-2\Lambda-6\varrho\right\},\label{con1}\\
H^i&=&-2\pi^{ij}_{~;j}=-2\pi^{ij}_{~,j}-h^{il}\left(2h_{jl,k}-h_{jk,l}\right)\pi^{jk},\label{con2}\\
K_{ij}[h]&=&\dfrac{1}{2N}\left(N_{i|j}+N_{j|i}-\dot{h}_{ij}\right),\label{con0}
\end{eqnarray}
where (\ref{con0}) follows from the Gauss-Codazzi equations for embedded space. Here $K_{ij}$ is the extrinsic-curvature tensor, $\varrho$ is the stress-energy tensor projected onto unit timelike normal vector field
\begin{equation}
  n^\mu=\left[-\dfrac{1}{N},-\dfrac{N^i}{N}\right],~~n_{\mu}=\left[-N,0\right],~~n^{\mu}n_{\mu}=1,
\end{equation}
to induced embedded 3-space
\begin{equation}
  \varrho=n^\mu n^\nu T_{\mu\nu}=\dfrac{1}{N^2}T_{00}-\dfrac{N^i}{N^2}(T_{0i}+T_{i0})+\dfrac{N^iN^j}{N^2}T_{ij},
\end{equation}
and $\pi^{ij}$ is the canonical conjugate momentum field to the field $h_{ij}$
\begin{equation}\label{mom}
\pi^{ij}=\dfrac{\delta L}{\delta\dot{h}_{ij}}=\sqrt{h}\left(h^{ij}K^i_i[h]-K^{ij}[h]\right).
\end{equation}
Time-preservation requirement \cite{dir} of the primary constraints \cite{dew} for (\ref{gd})
\begin{eqnarray}
  \pi&=&\dfrac{\delta L}{\delta\dot{N}}\approx0,\\
  \pi^i&=&\dfrac{\delta L}{\delta\dot{N_i}}\approx0,
\end{eqnarray}
leads to the secondary constraints
\begin{eqnarray}
  H&\approx&0,\label{const}\\
  H^i&\approx&0,\label{const1}
\end{eqnarray}
called the Hamiltonian constraint and the diffeomorphism constraint, respectively. The diffeomorphism constraint (\ref{const1}) merely reflects spatial diffeoinvariance of the theory, and dynamics is given by the Hamiltonian constraint (\ref{const}). By using of the conjugate momentum field (\ref{mom}), the Hamiltonian constraint (\ref{const}) can be written in the equivalent form
\begin{equation}\label{con}
H=G_{ijkl}\pi^{ij}\pi^{kl}+\sqrt{h}\left(R[h]-2\Lambda-6\varrho\right)=0,
\end{equation}
called the Einstein--Hamilton--Jacobi equation \cite{ham1}--\cite{ham33}. Here
\begin{equation}
G_{ijkl}=\dfrac{1}{2}h^{-1/2}\left(h_{ik}h_{jl}+h_{il}h_{jk}-h_{ij}h_{kl}\right)
\end{equation}
is called the Wheeler superspace metric.
\subsection{The Wheeler--DeWitt equation}
The classical geometrodynamics is given by the Dirac--ADM Hamiltonian constraint (\ref{con}) and can be quantized by direct application of the Dirac primary quantization \cite{dir}
\begin{eqnarray}\label{dpq}
i\left[\pi^{ij}(x),h_{kl}(y)\right]&=&\dfrac{1}{2}\left(\delta_{k}^{i}\delta_{l}^{j}+\delta_{l}^{i}\delta_{k}^{j}\right)\delta^{(3)}(x,y),\\
i\left[\pi^i(x),N_j(y)\right]&=&\delta^i_j\delta^{(3)}(x,y),\\
i\left[\pi(x),N(y)\right]&=&\delta^{(3)}(x,y),
\end{eqnarray}
that in result demands to introduce the canonical conjugate momentum operator in the form
\begin{equation}
  {\pi}^{ij}=-i\dfrac{\delta}{\delta h_{ij}},~~{\pi}^j=-i\dfrac{\delta}{\delta N_j},~~{\pi}=-i\dfrac{\delta}{\delta N},
\end{equation}
and leads to the Wheeler--DeWitt equation \cite{dew}
\begin{equation}\label{wdw}
{H}\Psi[h_{ij}]=\left\{-G_{ijkl}\dfrac{\delta^2}{\delta h_{ij}\delta h_{kl}}+h^{1/2}\left(R[h]-2\Lambda-6\varrho\right)\right\}\Psi[h_{ij}]=0.
\end{equation}
Other first class constraints are conditions on the wave function $\Psi[h]$
\begin{equation}
  {\pi}\Psi[h_{ij}]=0,~~{\pi}^i\Psi[h_{ij}]=0,~~{H}^i\Psi[h_{ij}]=0,
\end{equation}
and the canonical commutation relations hold
\begin{equation}
\left[{\pi}(x),{\pi}^i(y)\right]=\left[{\pi}(x),{H}^i(y)\right]=\left[{\pi}^i(x),{H}^j(y)\right]=\left[{\pi}^i(x),{H}(y)\right]=0.
\end{equation}
In result $H_i$ are generators of diffeomorphisms $\widetilde{x}^i=x^i+\delta x^i$ \cite{dew}
\begin{eqnarray}
i\left[h_{ij},\int_{\partial M}H_{a}\delta x^a d^3x\right]&=&-h_{ij,k}\delta x^k-h_{kj}\delta x^{k}_{~,i}-h_{ik}\delta x^{k}_{~,j}~~,\\
i\left[\pi_{ij},\int_{\partial M}H_{a}\delta x^a d^3x\right]&=&-\left(\pi_{ij}\delta x^k\right)_{,k}+\pi_{kj}\delta x^{i}_{~,k}+\pi_{ik}\delta x^{j}_{~,k}~~,
\end{eqnarray}
which can be expressed also as constraints commutators
\begin{eqnarray}
  i\left[H_i(x),H_j(y)\right]\!\!&=&\!\!\int_{\partial M}H_{a}c^a_{ij}d^3z,\label{com1}\\
  i\left[H(x),H_i(y)\right]\!\!&=&\!\!H\delta^{(3)}_{,i}(x,y),\label{com2}\\
  i\left[\int_{\partial M}H\delta x_1d^3x,\int_{\partial M}H\delta x_2d^3x\right]\!\!&=&\!\!\int_{\partial M}H^a\left(\delta x_{1,a}\delta x_2-\delta x_1\delta x_{2,a}\right)d^3x,\label{com3}
\end{eqnarray}
where $H_i=h_{ij}H^j$, and $c^a_{ij}$'s are structure constants of diffeomorphism group
\begin{equation}
  c^a_{ij}=\delta^a_i\delta^b_j\delta^{(3)}_{,b}(x,z)\delta^{(3)}(y,z)-\delta^a_j\delta^b_i\delta^{(3)}_{,b}(y,z)\delta^{(3)}(x,z)
\end{equation}
Commutators (\ref{com1}-\ref{com3}) show the first-class constrained system property.

\newpage
\section{Bosonization. Global One-Dimensionality.}
Commonly the Wheeler--DeWitt theory (\ref{wdw}) is interpreted in terms of the nonrelativistic Schr\"odinger quantum mechanics in configuration space of General Relativity. This point of view seems to be misleading, the conception of superspace is rather mysterious mathematical creation than real physical existence, and in this interpretation the Wheeler--Dewitt equation becomes physically senseless. Indeed one can ask: \emph{Why primary quantization of relativistic classical field theory, that is General Relativity, must be nonrelativistic Schr\"odinger quantum mechanics?} From conceptual point of view it is completely unnatural to interpret quantization of relativistic theory as nonrelativistic one. Really this question is old and seems to have answer in Dirac's considerations -- the result of classical field theory primary quantization should be relativistic quantum mechanics that is also a classical field theory. This is unique correct conceptual way on classical field theory level. However, in spite of this famous fact previous investigations of authors was concentrated on studying the Wheeler--DeWitt equation (\ref{wdw}) as a kind of nonrelativistic stationary wave mechanics. This quantum mechanical logics applied to quantization of the Einstein--Hilbert theory of gravitation is the most popular approach in the present state of quantum cosmology and quantum gravity (See, \emph{e.g.}, \cite{haw,abl,rov}). For example so called Loop Quantum Gravity and Loop Quantum Cosmology develop also quantum mechanics point of view. In result, in spite of beautiful philosophical as well as sophisticated mathematical constructions and many promises to gravitational physics, description of quantum gravity in terms of nonrelativistic quantum mechanics did not give any phenomenological results that can be confronted with experimental data.

By this reason in this section we will study the Wheeler--DeWitt equation (\ref{wdw}) from some new point of view, that is relativistic quantum mechanics as well as classical field theory. Recall that we have begun our considerations of quantum gravity by studying of $3+1$ decomposition of Lorentzian metric field of General Relativity, which means that actually we have considered some relativistic classical field theory after the Dirac primary quantization. From the famous candidates for the relativistic quantum mechanics equation it seems to be the best choice for this role the stationary the bosonic evolution, that is the Klein--Gordon--Fock wave equation. Indeed, the Wheeler--DeWitt theory is based on the second order differential equation in the superspace coordinate $h_{ij}$. However, it is some conceptual and formal problem to consider this equation with explicit presence of the superspace metrics $G_{ijkl}$. One can try to eliminate this refined tensor from our considerations and reduce $3+1$ Quantum Gravity to global one-dimensional classical field theory.

Let us consider the standard relation between functional differentials of 4-metric field and 4-volume form (See, \emph{e.g.}, \cite{mtw})
\begin{equation}\label{stan}
\delta g = gg^{\mu\nu}\delta g_{\mu\nu},
\end{equation}
where summation convention is assumed. By employing the $3+1$ decomposition (\ref{dec}) one can determine the variations of contravariant metric field components
\begin{eqnarray}
  \delta g_{00}&=&-\delta N^2+N^iN^j\delta h_{ij}+h_{ij}N^i\delta N^j+h_{ij}N^j\delta N^i,\\
  \delta g_{ij}&=&\delta h_{ij},\\
  \delta g_{0j}&=&h_{ij}\delta N^i+N^i\delta h_{ij},\\
  \delta g_{i0}&=&h_{ij}\delta N^j+N^j\delta h_{ij},
\end{eqnarray}
as well as the variation of 4-volume form
\begin{eqnarray}
\delta g=N^2\delta h+h\delta N^2.
\end{eqnarray}
So by using of covariant metric field components we obtain finally in result the formula
\begin{eqnarray}\label{hij}
  N^2\delta h=N^2hh^{ij}\delta h_{ij},
\end{eqnarray}
which establishes the global relation between 3-volume form and 3-metric field contravariant components. However, the relation (\ref{hij}) simultaneously allows to determine the functional derivative with respect to contravariant 3-metric field as an object proportional to covariant space metrics with functional differentiation with respect to the scalar field that is the space metrics determinant (3-volume form)
\begin{equation}
  \dfrac{\delta}{\delta h_{ij}} = hh^{ij}\dfrac{\delta}{\delta h}.
\end{equation}
Double using of this functional differential operator to the wave function of the Wheeler--DeWitt equation leads to
\begin{eqnarray}\label{bub}
  \dfrac{\delta}{\delta h_{ij}}\dfrac{\delta}{\delta h_{kl}}\Psi[h]&=&hh^{ij}\dfrac{\delta}{\delta h}\left(hh^{kl}\dfrac{\delta}{\delta h}\right)\Psi[h]=\nonumber\\
&=&hh^{ij}\left(h^{kl}\dfrac{\delta}{\delta h}+h\dfrac{\delta h^{kl}}{\delta h}\dfrac{\delta}{\delta h}+hh^{kl}\dfrac{\delta^2}{\delta h^2}\right)\Psi[h],
\end{eqnarray}
and by direct computation of the variation of covariant space metrics with using by the relation (\ref{stan})
\begin{equation}
\delta h^{ij}=\delta\dfrac{1}{h_{ij}}=-\dfrac{\delta h_{ij}}{\left(h_{ij}\right)^2},
\end{equation}
we obtain as result
\begin{equation}
  \delta h^{ij}=-\dfrac{h^{ij}}{h}\delta h\longrightarrow\dfrac{\delta h^{ij}}{\delta h}=-\dfrac{h^{ij}}{h},
\end{equation}
that after application in the second term of (\ref{bub}) causes that two first terms vanishes and in result we obtain important for further development conclusion
\begin{equation}\label{key}
  \dfrac{\delta}{\delta h_{ij}}\dfrac{\delta}{\delta h_{kl}}\Psi[h]=h^2h^{ij}h^{kl}\dfrac{\delta^2\Psi[h]}{\delta h^2}.
\end{equation}
One can see now that really we have not to deal with the superspace metrics $G_{ijkl}$ explicitly, namely the key relation (\ref{key}) in result leads to the scalar beeing double contraction of the superspace metrics
\begin{eqnarray}
  G_{ijkl}h^{ij}h^{kl}=\dfrac{1}{2}h^{-1/2}\left(h_{ik}h_{jl}+h_{il}h_{jk}-h_{ij}h_{kl}\right)h^{ij}h^{kl}=-\dfrac{3}{2}h^{-1/2},
\end{eqnarray}
and by this the functional differentiation with respect to space metrics being an origin of the Wheeler--DeWitt equation transits into the functional differentiation with respect to space metrics determinant with some scalar coefficient
\begin{eqnarray}
  -G_{ijkl}\dfrac{\delta^2\Psi[h]}{\delta h_{ij}\delta h_{kl}}=\dfrac{3}{2}h^{3/2}\dfrac{\delta^2\Psi[h]}{\delta h^2}.
\end{eqnarray}
In consequence we conclude that the superspatial Wheeler--DeWitt equation (\ref{wdw}) transforms by the following way
\begin{equation}
  \left\{\dfrac{3}{2}h^{3/2}\dfrac{\delta^2}{\delta h^2}+h^{1/2}\left(R[h]-2\Lambda-6\varrho\right)\right\}\Psi[h]=0.
\end{equation}
This equation can be rewritten in form of the functional 1--dimensional Klein--Gordon--Fock equation for the classical massive Bose field $\Psi[h]$
\begin{equation}\label{new}
\left\{\dfrac{\delta^2}{\delta{h^2}}+m^2[h]\right\}\Psi[h]=0,
\end{equation}
that lies in accordance with general relativistic character of the classical Einstein--Hilbert theory of gravitation. Formally one can understand the quantity
\begin{equation}\label{masqr}
\dfrac{2}{3h}\left(R[h]-2\Lambda-6\varrho\right)\equiv m^2[h],
\end{equation}
as square of mass for the classical bosonic field $\Psi[h]$, and build a quantum theory of the Einstein--Hilbert general gravitational fields as quantum field theory, where classical Riemannian manifold is an effect of the Bose gas.

\section{Space quantum states}
The previous section was devoted to presentation of results for the quantum geometrodynamics treated in terms of the relativistic quantum mechanics defined by the Klein--Gordon--Fock equation (\ref{kgf}) for the classical Bose field $\Psi[h]$ associated with the Einstein--Hilbert Riemannian manifold of General Relativity. In the present section we will construct quantum field theory of the considered Bose field by language of the Fock space of annihilation and creation operators.
\subsection{Canonical reduction}
Let us consider again the Klein--Gordon--Fock equation (\ref{kgf}). Formally one can consider this Euler--Lagrange equation of motion as combination of two equations: the equation of motion for the canonical conjugate momentum field given by (\ref{lagr1}) and the constraint for the canonical conjugate momentum field that is (\ref{lagr2}). This system of equations can be rewritten in the reduced form, that is the first order functional differential equation as follows
\begin{equation}\label{red}
  \dfrac{\delta}{\delta{h}}\left[\begin{array}{c}\Psi\\\Pi_\Psi\end{array}\right]=\left[\begin{array}{cc}0&1\\-m^2[h]&0\end{array}\right]\left[\begin{array}{c}\Psi\\\Pi_\Psi\end{array}\right].
\end{equation}
With using of the following abbreviated notation
\begin{equation}
  \Phi_{\mu}=\left[\begin{array}{c}\Psi\\ \Pi_\Psi\end{array}\right],~~\partial_{\nu}=\left[\begin{array}{c}\dfrac{\delta}{\delta h}\\0\end{array}\right],
\end{equation}
the reduced equation (\ref{red}) can be presented in the form that is looks like formally to the Dirac equation
\begin{equation}\label{dir}
 \left(i\Gamma^{\mu}\partial_{\nu}-\mathrm{M}^{\mu}_{\nu}\right)\Phi_{\mu}=0,
\end{equation}
where the positively defined mass matrix $\mathrm{M}^{\mu}_{\nu}$ is determined by
\begin{equation}\label{masmat}
  \mathrm{M}^{\mu}_{\nu}=\left[\begin{array}{cc}
0&-1\\-m^{2}&0\end{array}\right]\geq0.
\end{equation}
However, in the considered case the matrices $\Gamma^{\mu}=\left[-i\mathbf{I}_2,\mathbf{0}_{2}\right]$ create different the Clifford algebra than in the case the Dirac algebra
\begin{equation}
  \left\{\Gamma^{\mu},\Gamma^{\nu}\right\}=2\eta^{\mu\nu}\mathbf{I}_2,
\end{equation}
where $\{,\}$ are anticommutator brackets, $\mathbf{I}_2$ is 2-dimensional unit matrix, and $\mathbf{0}_2$ is 2-dimensional null matrix, and the metrics $\eta_{\mu\nu}$ in this case is given by
\begin{equation}
  \eta^{\mu\nu}=\left[\begin{array}{cc}-1&0\\0&0\end{array}\right],
\end{equation}
that is agreed with 1-dimensional equation (\ref{kgf}).

Let us investigate the generalized second quantization of the reduced relativistic quantum mechanics (\ref{dir}) by application of the Fock space of creation and annihilation functional operators. Note that the classical field theory Hamiltonian of considered system given by the relation (\ref{ham})
\begin{equation}\label{ham0}
H[h]=\dfrac{\Pi_\Psi^2[h]+m^2[h]\Psi^2[h]}{2},
\end{equation}
can be presented in the matrix form
\begin{equation}\label{ham1}
  H[h]=[\Psi, \Pi_\Psi]\left[\begin{array}{cc}\alpha&\beta\\\gamma&\delta\end{array}\right]\left[\begin{array}{c}\Psi\\ \Pi_\Psi\end{array}\right]=\Phi_{\mu}^\dagger H^{\mu\nu}\Phi_{\nu},
\end{equation}
where $\alpha,~\beta,~\gamma,~\delta$ are some functionals of $h$, generally. From classical point of view the Hamiltonian (\ref{ham1}) can be rewritten as
\begin{equation}\label{ham2}
  H[h]=\alpha\Psi^2+\delta\Pi_\Psi^2+\gamma\Pi_\Psi\Psi+\beta\Psi\Pi_\Psi,
\end{equation}
with natural identification concluded from the form of the Hamiltonian (\ref{ham0})
\begin{equation}
  \alpha=\dfrac{1}{2},~~\delta=\dfrac{1}{2}m^2[h],~~\beta=\gamma=0.
\end{equation}
Let us build the second quantization of the reduced equation (\ref{dir}) based on quantization of the classical field theory Hamiltonian.
\subsection{Second quantization in the Fock space}
The quantization of the considered classical Bose field theory will understand in this paper in terms of the fundamental operator quantization of the reduced Klein--Gordon--Fock field equation (\ref{dir}). This quantization, called the second quantization, that can be presented formally as the transition between classical fields and quantum fields operators as follows
\begin{eqnarray}
\left[\begin{array}{c}\Psi\\ \Pi_\Psi\end{array}\right]\longrightarrow\left[\begin{array}{c}\mathbf{\Psi}\\\mathbf{\Pi}_\Psi\end{array}\right],
\end{eqnarray}
and applied to the reduced equation (\ref{dir}) gives as the result the following functional operator equation
\begin{equation}
  \left(i\Gamma^{\mu}\partial_{\nu}-\mathrm{M}^{\mu}_{\nu}\right)\mathbf{\Phi}_{\mu}=0.\label{qkg}
\end{equation}
According to standard rules of quantum field theory, the quantization must be constructed by application of canonical commutation relations that are agreed with quantum statistics that have the entrance relativistic quantum mechanical equation given in the considered case by the Klein--Gordon--Fock equation (\ref{kgf}). Obviously, this is the Bose statistics, and in this manner we should apply the standard rules of bosonic quantum field theory thats are \cite{bog}
\begin{eqnarray}
  \left[\mathbf{\Pi}_{\Psi}[h'],\mathbf{\Psi}[h]\right]&=&-i\delta(h'-h),\label{c1}\\
  \left[\mathbf{\Pi}_{\Psi}[h'],\mathbf{\Pi}_{\Psi}[h]\right]&=&0,\label{c2}\\
  \left[\mathbf{\Psi}[h'],\mathbf{\Psi}[h]\right]&=&0.\label{c3}
\end{eqnarray}
where $[,]$ are commutator brackets. From the quantum field theory point of view, that we want to construct here, the classical field theory Hamiltonian (\ref{ham0}) must be quantized in terms of field operators, and in this case we should consider instead the classical Hamiltonian (\ref{ham1}) more general quadratic form
\begin{eqnarray}
  \mathbf{H}[h',h]&=&[\mathbf{\Psi}[h'], \mathbf{\Pi}_\Psi[h']]\left[\begin{array}{cc}\alpha[h',h]&\beta[h',h]\\
  \gamma[h',h]&\delta[h',h]\end{array}\right]\left[\begin{array}{c}\mathbf{\Psi}[h]\\ \mathbf{\Pi}_\Psi[h]\end{array}\right]\equiv\nonumber\\
  &\equiv&\mathbf{\Phi}_{\mu}^\dagger[h'] H^{\mu\nu}[h',h]\mathbf{\Phi}_{\nu}[h],
\end{eqnarray}
and the formula (\ref{ham2}) in this case has a form
\begin{eqnarray}\label{hamx0}
  \mathbf{H}[h',h]&=&\alpha[h',h]\mathbf{\Psi}[h']\mathbf{\Psi}[h]+\delta[h',h]\mathbf{\Pi}_\Psi[h']\mathbf{\Pi}_\Psi[h]+\nonumber\\
  &+&\gamma[h',h]\mathbf{\Pi}_\Psi[h']\mathbf{\Psi}[h]+\beta[h',h]\mathbf{\Psi}[h']\mathbf{\Pi}_\Psi[h],
\end{eqnarray}
that by existence of the canonical commutation relations (\ref{c1}), (\ref{c2}), (\ref{c3}) is nonequivalent to the classical field theory Hamiltonian (\ref{ham1}).
Now we want to see directly that if we want to preserve in quantum field theory the classical form of the field Hamiltonian (\ref{ham1}), \emph{i.e.}
\begin{equation}\label{hamx}
\mathbf{H}[h]=\dfrac{\mathbf{\Pi}_\Psi^2[h]+m^2[h]\mathbf{\Psi}^2[h]}{2},
\end{equation}
where $\mathbf{H}[h]\equiv\mathbf{H}[h,h]$, then we must take into consideration the following identification
\begin{equation}
\alpha[h',h]=\dfrac{1}{2},~~\delta[h',h]=\dfrac{1}{2}m[h']m[h],~~\gamma[h',h]=C,~~\beta[h',h]=-C,
\end{equation}
where $C$ is some constant c-number independent on $h$. Then from (\ref{hamx0}) we obtain directly
\begin{equation}\label{hf}
  \mathbf{H}[h]=\dfrac{\mathbf{\Pi}_\Psi^2[h]+m^2[h]\mathbf{\Psi}^2[h]}{2}-iC\delta(0),
\end{equation}
where the last term can be omitted by c-number character. However, generally the quantum field theory of considered boson has the following field Hamiltonian
\begin{equation}\label{hff}
  \mathbf{H}[h]=\mathbf{\Phi}^\dagger[h]\left[\begin{array}{cc}\dfrac{1}{2}&-C\\
  C&\dfrac{m^2[h]}{2}\end{array}\right]\mathbf{\Phi}[h],
\end{equation}
that is obviously nondiagonal. Because of, as it was seen in the relation (\ref{hf}), the constant c-number $C$ does not play role in physics - it is only a choice of reference Hamiltonian value - one can put into computations the simplest case $C\equiv0$. Then the quantum field theory Hamiltonian (\ref{hff}) is diagonal, and moreover the demanded classical form of the quantum field Hamiltonian is preserved
\begin{equation}\label{hf1}
  \mathbf{H}[h]=\dfrac{\mathbf{\Pi}_\Psi^2[h]+m^2[h]\mathbf{\Psi}^2[h]}{2}.
\end{equation}
In this manner we must not search for special diagonalizable basis, and we can directly apply to the system the Fock space quantization.

We propose apply to the reduced Klein--Gordon--Fock equation (\ref{qkg}) the following generalized fundamental operator quantization in the Fock space of creation and annihilation functional operators $\mathsf{G}^\dagger[h]$ and $\mathsf{G}[h]$
\begin{eqnarray}\label{sq}
\left[\begin{array}{c}\mathbf{\Psi}[h]\\ \mathbf{\Pi}_\Psi[h]\end{array}\right]
=\left[\begin{array}{cc}\dfrac{1}{\sqrt{2|m[h]|}}&\dfrac{1}{\sqrt{2|m[h]|}}\\
-i\sqrt{\dfrac{|m[h]|}{2}}&i\sqrt{\dfrac{|m[h]|}{2}}\end{array}\right]
\left[\begin{array}{c}\mathsf{G}[h]\\ \mathsf{G}^{\dagger}[h]\end{array}\right],
\end{eqnarray}
Let us note that this second quantization lies in strict accordance with the bosonic character of the equation (\ref{kgf}), and also with the generalized algebraic approach investigated in papers of Von Neumann, Araki and Woods \cite{neu, aw}. The principal canonical commutation relations (\ref{c1}), (\ref{c2}), and (\ref{c3}) are automatically fulfilled if the considered Bose system is described by the dynamical basis $\mathfrak{B}[h]$ in the proposed Fock space construction
\begin{equation}\label{db}
  \mathfrak{B}[h]=\left\{\left[\begin{array}{c}\mathsf{G}[h]\\
\mathsf{G}^{\dagger}[h]\end{array}\right]:\left[\mathsf{G}[h'],\mathsf{G}^{\dagger}[h]\right]=\delta\left(h'-h\right), \left[\mathsf{G}[h'],\mathsf{G}[h]\right]=0\right\},
\end{equation}
so that the Fock space quantization (\ref{sq}) can be rewritten briefly as action of the second quantization matrix on the dynamical basis
\begin{equation}\label{sqx}
  \mathbf{\Phi}[h]=\mathbb{Q}[h]\mathfrak{B}[h],
\end{equation}
where the second quantization matrix can be determined directly as
\begin{equation}\label{sqxm}
  \mathbb{Q}[h]=\left[\begin{array}{cc}\dfrac{1}{\sqrt{2|m[h]|}}&\dfrac{1}{\sqrt{2|m[h]|}}\\
-i\sqrt{\dfrac{|m[h]|}{2}}&i\sqrt{\dfrac{|m[h]|}{2}}\end{array}\right].
\end{equation}
The quantum field dynamics considered from the point of view of the basis $\mathfrak{B}[h]$ is described by the Klein--Gordon--Fock equation (\ref{kgf}) with application of the second quantization (\ref{sqx}). By elementary calculations one can obtain that the $h$-evolution of the basis $\mathfrak{B}[h]$ is governed by the following equation of motion
\begin{equation}\label{df}
\dfrac{\delta\mathfrak{B}[h]}{\delta h}=\left[\begin{array}{cc}
-im[h]&\dfrac{1}{2m[h]}\dfrac{\delta m[h]}{\delta h}\\
\dfrac{1}{2m[h]}\dfrac{\delta m[h]}{\delta h}&im[h]\end{array}\right]\mathfrak{B}[h].
\end{equation}
Formally, this is nonlinear first order functional differential equation. This system of equations for the creation and annihilation functional operators can not be solved directly by application of the standard path integrals method, here the coupling between annihilation and creation operators is present in form of nondiagonal elements. Moreover, in the dynamical basis (\ref{db}) we have to deal with some kind of global situation -- by reference frame dependence of the quantum vacuum state, vacuum expectation values of the quantum field theory Hamiltonian (\ref{hf1}) that in the dynamical basis has a following form
\begin{eqnarray}\label{qfth}
\mathbf{H}[h]&=&\mathfrak{B}^\dagger[h]\left[\begin{array}{cc}\dfrac{m[h]}{2}&0\\0&\dfrac{m[h]}{2}\end{array}\right]\mathfrak{B}[h]=\nonumber\\
&=&m[h]\left(\mathsf{G}^\dagger[h]\mathsf{G}[h]+\dfrac{1}{2}\delta(0)\right),
\end{eqnarray}
can not be treated as correctly defined, by the dynamical character of the basis. For correctness we must build some static functional operator basis -- in this type of basis, the vacuum expectation values can be determined by local status of the basis, and by this quantum field theory is no senseless. The recept to the similar evolutions is only one, unique, and unambiguous -- this is diagonalization of the operator evolution to the Heisenberg canonical form together with using of the Bogoliubov transformation agreed with canonical commutation relations.

Let us apply the diagonalization procedure. Firstly, we take into considerations the supposition that sounds that in the Fock space exists some local basis where exactly the same canonical commutation relations between creation and annihilation functional operators are preserved
\begin{equation}
  \mathfrak{B}^\prime[h]=\left\{\left[\begin{array}{c}\mathsf{G}^\prime[h]\\
\mathsf{G}^{\prime\dagger}[h]\end{array}\right]:\left[\mathsf{G}^\prime[h'],\mathsf{G}^{\prime\dagger}[h]\right]=\delta\left(h'-h\right), \left[\mathsf{G}^\prime[h'],\mathsf{G}^\prime[h]\right]=0\right\}.
\end{equation}
This basis in our studies has the fundamental status, namely we suppose that in this basis the functional operator evolution (\ref{df}) diagonalizes directly to the canonical Heisenberg operator evolution
\begin{equation}
\dfrac{\delta\mathfrak{B}^\prime[h]}{\delta h}=\left[\begin{array}{cc}
-i\lambda[h] & 0 \\ 0 &
i\lambda[h]\end{array}\right]\mathfrak{B}^\prime[h],
\end{equation}
where $\lambda[h]$ is generally some functional of the evolution parameter. By using of the supposition that in the local basis are preserved the bosonic canonical commutation relations, one can deduce directly that the fundamental basis $\mathfrak{B}^\prime[h]$ should be obtained from the dynamical basis $\mathfrak{B}[h]$ by some generalized canonical transformation in the considered Fock space, that is a rotation of basis determined by standardly defined the Bogoliubov transformation, which in the case of systems with Bose statistics has the following form
\begin{equation}
\mathfrak{B}^\prime[h]=\left[\begin{array}{cc}u[h]&v[h]\\
v^{\ast}[h]&u^{\ast}[h]\end{array}\right]\mathfrak{B}[h],
\end{equation}
where the Bogoliubov coefficients $u$ and $v$ are functionals of $h$, and by rotational character of the bosonic Bogoliubov transformation these coefficients obey the Lobachevskiy--Gauss--Bolyai hyperbolic space condition
\begin{equation}\label{lgb}
  |u[h]|^2-|v[h]|^2=1.
\end{equation}
This can be checked by direct elementary computation that the proposed two-step diagonalization operator evolution equations procedure in result leads to demanding of vanishing of the functional $\lambda[h]$, but simultaneously transits a whole dynamical evolution from the operator basis onto the system of the bosonic Bogoliubov coefficients
\begin{equation}\label{bcof}
  \dfrac{\delta}{\delta h}\left[\begin{array}{c}u[h]\\v[h]\end{array}\right]=\left[\begin{array}{cc}
-im[h]&\dfrac{1}{2m[h]}\dfrac{\delta m[h]}{\delta h}\\
\dfrac{1}{2m[h]}\dfrac{\delta m[h]}{\delta h}&im[h]\end{array}\right]\left[\begin{array}{c}u[h]\\v[h]\end{array}\right].
\end{equation}
By this the procedure leads to realization of the main aim of this construction - namely this gives definition of the static operator basis that has the fundamental character; the static basis is completely determined by initial data problem in the Fock space, and is given by usual ladder operators
\begin{equation}\label{in}
\mathfrak{B}^\prime[h]=\mathfrak{B}_{I}=\left\{\left[\begin{array}{c}\mathsf{G}_I\\
\mathsf{G}^{\dagger}_I\end{array}\right]: \left[\mathsf{G}_I,\mathsf{G}^{\dagger}_I\right]=1, \left[\mathsf{G}_I,\mathsf{G}_I\right]=0\right\}.
 \end{equation}
Furthermore, by the static character the fundamental operator basis, this basis defines static quantum vacuum state given as
\begin{equation}
|0\rangle_I=\left\{|0\rangle_I:\mathsf{G}_I|0\rangle_I=0,~0={_I}\langle0| \mathsf{G}_I^\dagger\right\},
\end{equation}
and vacuum expectation values computed on this \emph{initial data} vacuum state are well-defined by local status of the fundamental basis $\mathfrak{B}_{I}$.

The functional differential equations for the Bogoliubov coefficients (\ref{bcof}) can be solved directly by famous methods of linear analytical algebra based on the Cayley--Hamilton theorem. However, in the present situation we have very special evolution - actually the bosonic Bogoliubov coefficients can not be chosen arbitrary, by the fact that they are constrained by the rotational condition (\ref{lgb}), and in possible solving method we should construct firstly some parametrization that lies in accordance with this hyperbolic constraint, and then to try solve the coefficients evolution equation (\ref{bcof}) in this concretely chosen parametrization. It can be checked by direct algebraic manipulation that reverse conduct leads to bad-defined algebraical problem. By hyperbolic view of the rotational condition (\ref{lgb}) we suggest to use the very special parametrization of the bosonic Bogoliubov coefficients, so called the superfluid coordinate system defined by the following transformation
\begin{eqnarray}\label{sup}
u[h]&=&\exp\left\{i\theta[h]\right\}\cosh \phi[h],\\
v[h]&=&\exp\left\{i\theta[h]\right\}\sinh \phi[h].
\end{eqnarray}
Elementary algebraic manipulations lead to the system of functional differential equations for the parameters $\theta[h]$ and $\phi[h]$, that can be solved directly and the solutions can be written in the following form
\begin{eqnarray}
\theta[h]&=&\pm m_I\int_{h_I}^{h}\sqrt{\left|\dfrac{m^2[h]}{m_I^2}\right|}\delta h,\label{phas}\\
\phi[h]&=&\ln{\sqrt[4]{\left|\dfrac{m^2[h]}{m_I^2}\right|}},~~m_I=m[h_I].
\end{eqnarray}
The interpretation of these solutions is obvious -- the quantity $\theta[h]$ is integrated mass of the considered boson, and the solution $\phi[h]$ is logarithmic field associated with mass of the boson. By this the bosonic Bogoliubov coefficients (\ref{bcof}) can be determined as follows
\begin{eqnarray}
u[h]&=&\dfrac{1}{2}\exp\left\{\pm im_I\int_{h_I}^{h}\dfrac{m[h]}{m_I}\delta h\right\}\left(\sqrt[4]{\left|\dfrac{m^2[h]}{m_I^2}\right|}+\sqrt[4]{\left|\dfrac{m_I^2}{m^2[h]}\right|}\right),\label{sup1}\\
v[h]&=&\dfrac{1}{2}\exp\left\{\pm im_I\int_{h_I}^{h}\dfrac{m[h]}{m_I}\delta h\right\}\left(\sqrt[4]{\left|\dfrac{m^2[h]}{m_I^2}\right|}-\sqrt[4]{\left|\dfrac{m_I^2}{m^2[h]}\right|}\right).\label{sup2}
\end{eqnarray}
We have a freedom in sign choosing of the phases, but we decide to choose the positive phases. Actually by definition of the coefficients $u$ and $v$, we determinate the monodromy matrix $\mathbb{G}[h]$ that transits the fundamental basis $\mathfrak{B}_I$ into the dynamical one $\mathfrak{B}[h]$ defined as
\begin{equation}
  \mathfrak{B}[h]=\mathbb{G}[h]\mathfrak{B}_I
\end{equation}
that has the following form
\begin{eqnarray}\label{mon}
\mathbb{G}[h]\!=\!\left[\!\!\!\begin{array}{cc}
\left(\sqrt[4]{\left|\dfrac{m_I^2}{m^2[h]}\right|}\!+\!\sqrt[4]{\left|\dfrac{m^2[h]}{m_I^2}\right|}\right)\dfrac{e^{-i\theta[h]}}{2}\vspace*{10pt}\!\!\!\!&
\left(\sqrt[4]{\left|\dfrac{m_I^2}{m^2[h]}\right|}\!-\!\sqrt[4]{\left|\dfrac{m^2[h]}{m_I^2}\right|}\right)\dfrac{e^{i\theta[h]}}{2}\\
\left(\sqrt[4]{\left|\dfrac{m_I^2}{m^2[h]}\right|}\!-\!\sqrt[4]{\left|\dfrac{m^2[h]}{m_I^2}\right|}\right)\dfrac{e^{-i\theta[h]}}{2}\!\!\!\!&
\left(\sqrt[4]{\left|\dfrac{m_I^2}{m^2[h]}\right|}\!+\!\sqrt[4]{\left|\dfrac{m^2[h]}{m_I^2}\right|}\right)\dfrac{e^{i\theta[h]}}{2}\end{array}\!\!\!\right]\!\!.
\end{eqnarray}
In this manner one can conclude directly that in the presented approach the quantum theory of gravitation is completely determined by the correct choose of the monodromy matrix between dynamic and static bases in the Fock space of creation and annihilation functional operators. By this reason in the Fock space formulation, quantum gravitation one can immediately understood as the phenomena that is an effect of the choice of operator basis. The initial data basis $\mathfrak{B}_I$ is directly related with intial data of the creation and annihilation operators, and in this manner this has the fundamental status for description of physical phenomena -- the monodromy matrix (\ref{mon}) consists whole information about dynamics of a space geometry of the Riemannian manifold given by a solution of the Einstein--Hilbert field equations of General Relativity (\ref{gr1}). Moreover, one can see directly from the form of the monodromy matrix (\ref{mon}), that this fundamental quantity is completely determined by a quotient of two squares of mass for the Bose field - one taken in the initial point, and the second taken in the current evolution point. By this reason, actually this quotient of squares of mass has the fundamental physical meaning for the quantum theory of the considered Bose field.
%\pagebreak
\subsection{Quantum bosonic field. One--point correlations.}
The field operator $\mathbf{\Phi}[h]$ which is directly associated with a 3-dimensional spatial part of the Einstein--Hilbert Riemannian manifold of General Relativity, and represents a spacetime in terms of bosonic quantum field theory can be concluded immediately as an effect of transformation of the fundamental static initial data basis by directed action of the monodromy matrix $\mathbb{G}[h]$ and the second quantization matrix $\mathbb{Q}[h]$ as follows
\begin{equation}
  \mathbf{\Phi}[h]=\mathbb{Q}[h]\mathbb{G}[h]\mathfrak{B}_I.
\end{equation}
In this manner by multiplication of matrices $\mathbb{Q}$ and $\mathbb{G}$ given by the formulas (\ref{sqx}) and (\ref{mon}) the bosonic field operator associated with a spatial geometry of spacetime can be concluded in the form
\begin{eqnarray}\label{field}
  \mathbf{\Psi}[h]=\frac{1}{2\sqrt{2m_I}}\sqrt{\dfrac{m_I^2}{m^2[h]}}\left(e^{-i\theta[h]}\mathsf{G}_I+e^{i\theta[h]}\mathsf{G}^{\dagger}_I\right).
\end{eqnarray}
This field operator is formally hermitian operator
\begin{eqnarray}
  \mathbf{\Psi}^\dagger[h]=\mathbf{\Psi}[h],
\end{eqnarray}
and acts on the static vacuum state by the following way
\begin{eqnarray}
  \mathbf{\Psi}[h]|0\rangle_I&=&\frac{1}{2\sqrt{2m_I}}\sqrt{\dfrac{m_I^2}{m^2[h]}}e^{i\theta[h]}\mathsf{G}^{\dagger}_I|0\rangle_I,\\
  {_I}\langle0|\mathbf{\Psi}^\dagger[h]&=&{_I}\langle0|\mathsf{G}_I\frac{1}{2\sqrt{2m_I}}\sqrt{\dfrac{m_I^2}{m^2[h]}}e^{-i\theta[h]}.
\end{eqnarray}
Linear algebra gives the theorem that states that eigenvalues of operator function are given by functions of the operator eigenvalues, and in considered case it particularly allows to define the multifield quantum states
\begin{eqnarray}
\left(\mathbf{\Psi}[h]\right)^n|0\rangle_I&=&\left(\frac{1}{2\sqrt{2m_I}}\sqrt{\dfrac{m_I^2}{m^2[h]}}e^{i\theta[h]}\right)^n\mathsf{G}^{\dagger n}_I|0\rangle_I,\\
  {_I}\langle0|\left(\mathbf{\Psi}^\dagger[h']\right)^{n'}&=&{_I}\langle0|\mathsf{G}_I^{n'}\left(\frac{1}{2\sqrt{2m_I}}\sqrt{\dfrac{m_I^2}{m^2[h]}}e^{-i\theta[h']}\right)^{n'},
\end{eqnarray}
where $n'$ and $n$ are natural numbers, and $h'$ and $h$ are determinants of space metrics and characterize quantum state of spacetime given by a metrics with space part described respectively by $h_{\mu\nu}^\prime$ and $h_{\mu\nu}$. We will these states as \emph{space quantum states of a spacetime} and for shortness we will note these states as
\begin{eqnarray}
  \left(\mathbf{\Psi}[h]\right)^n|0\rangle_I&\equiv&|h,n\rangle,\\
    {_I}\langle0|\left(\mathbf{\Psi}^\dagger[h']\right)^{n'}&\equiv&\langle n',h'|.
\end{eqnarray}
Generalized two-point correlation functions of two space quantum states can be determined immediately as
\begin{eqnarray}\label{gencor}
  \langle n',h'|h,n\rangle=\dfrac{m_I^{(n'+n)/2}}{2^{3(n'+n)/2}}\frac{e^{-i(n'\theta[h']-n\theta[h])}}{(m[h'])^{n'}(m[h])^n}{_I}\langle0|\mathsf{G}_I^{n'}\mathsf{G}^{\dagger n}_I|0\rangle_I.\label{cor0}
\end{eqnarray}
By normalization to unity the initial data correlator
\begin{equation}
  \langle 1,h_I|h_I,1\rangle=\dfrac{1}{8m_I}{_I}\langle0|0\rangle_I\equiv1,
\end{equation}
one can determinate the vacuum-vacuum amplitude as
\begin{equation}
  {_I}\langle0|0\rangle_I=8m_I,
\end{equation}
and simultaneously it can be treated as the definition of initial data mass $m_I$. Especially interesting for further developments correlators are
\begin{eqnarray}
  \langle 1,h|h,1\rangle&=&\dfrac{m_I^2}{m^2[h]},\label{cor2}\\
  \langle n',h|h,n\rangle&=&\left(\dfrac{\langle 1,h|h,1\rangle}{{_I}\langle0|0\rangle_I}\right)^{(n'+n)/2}e^{-i(n'-n)\theta[h]}{_I}\langle0|\mathsf{G}_I^{n'}\mathsf{G}^{\dagger n}_I|0\rangle_I,\label{cor4}\\
  \langle 1,h'|h,1\rangle&=&\dfrac{m_I^2}{m[h']m[h]}\exp\left\{i\int_{h'}^{h}m[h'']\delta h''\right\},\label{cor1}\\
  \dfrac{\langle n,h'|h,n\rangle}{{_I}\langle 0|0\rangle_I}&=&\left(\dfrac{\langle 1,h'|h,1\rangle}{{_I}\langle0|0\rangle_I}\right)^n,\label{cor3}
\end{eqnarray}
where in calculations of vacuum expectation values was used the identity
\begin{eqnarray}
  {_I}\langle0|\mathsf{G}_I^{n}\mathsf{G}^{\dagger n}_I|0\rangle_I&=&{_I}\langle0|\left(\mathsf{G}_I\mathsf{G}^{\dagger}_I\right)^n|0\rangle_I={_I}\langle0|\left(1+\mathsf{G}^{\dagger}_I\mathsf{G}_I\right)^n|0\rangle_I=\nonumber\\
  &=&{_I}\langle0|\sum_{k=0}^nC^n_k\left(\mathsf{G}^{\dagger}_I\mathsf{G}_I\right)^k|0\rangle_I=\sum_{k=0}^nC^n_k{_I}\langle0|\left(\mathsf{G}^{\dagger}_I\mathsf{G}_I\right)^k|0\rangle_I=\nonumber\\
  &=&C^n_0{_I}\langle0|0\rangle_I=8m_I,
\end{eqnarray}
with $C^n_k=\dfrac{n!}{k!(n-k)!}$ being the Newton binomial coefficient. The correlator (\ref{cor2}) is basic, naturally one can find from (\ref{gencor}) that
\begin{equation}
  \dfrac{\langle n',h'|h,n\rangle}{{_I}\langle0|0\rangle_I^{(n'+n)/2}}=\sqrt{\langle 1,h'|h',1\rangle^{n'}\langle 1,h|h,1\rangle^n}e^{-im_I\theta_{n',n}[h',h]}{_I}\langle0|\mathsf{G}_I^{n'}\mathsf{G}^{\dagger n}_I|0\rangle_I,
\end{equation}
where
\begin{equation}
\theta_{n',n}[h',h]=n'\int_{h_I}^{h'}\dfrac{\delta h''}{\sqrt{\langle 1,h''|h'',1\rangle}}-n\int_{h_I}^{h}\dfrac{\delta h''}{\sqrt{\langle 1,h''|h'',1\rangle}}.
\end{equation}
For example one can directly define the two-point correlator (\ref{cor1}) in the following way
\begin{eqnarray}\label{2p}
  \langle 1,h'|h,1\rangle=\sqrt{\langle 1,h'|h',1\rangle\langle 1,h|h,1\rangle}\exp\left\{im_I\int_{h'}^{h}\dfrac{\delta h''}{\sqrt{\langle 1,h''|h'',1\rangle}}\right\},
\end{eqnarray}
or by application of the functional Taylor series expansion of the integral in exponent of the last relation
\begin{equation}
\int_{h'}^{h}\dfrac{\delta h''}{\sqrt{\langle 1,h''|h'',1\rangle}}=\sum_{n=0}^{\infty}\kappa_n[h,h'|h_I]{\dfrac{\delta^n}{\delta h^n}\langle 1,h|h,1\rangle\Biggr|_{h_I}},
\end{equation}
with the functional coefficients
\begin{equation}
  \kappa_n[h,h'|h_I]=\dfrac{(2n-3)!}{2^{2n-1}(n-1)!}\sum_{k=0}^{n+1}\dfrac{(-1)^k}{k!(n-k+1)!}(h_I)^{n-k+1}\left(h^k-h'^k\right),
\end{equation}
the considered two-point correlator becomes
\begin{equation}
  \langle 1,h'|h,1\rangle=\sqrt{\langle 1,h'|h',1\rangle\langle 1,h|h,1\rangle}\prod_{n=0}^{\infty}\sum_{p=0}^{\infty}\dfrac{\left(\kappa_n^\prime\right)^p}{p!}\left(\dfrac{\delta^n}{\delta h^n}\langle 1,h|h,1\rangle\Biggr|_{h_I}\right)^p,
\end{equation}
where for shortness $\kappa_n^\prime\equiv im_I\kappa_n[h,h'|h_I]$.

\begin{figure}
  \centering
  \includegraphics[scale=0.5]{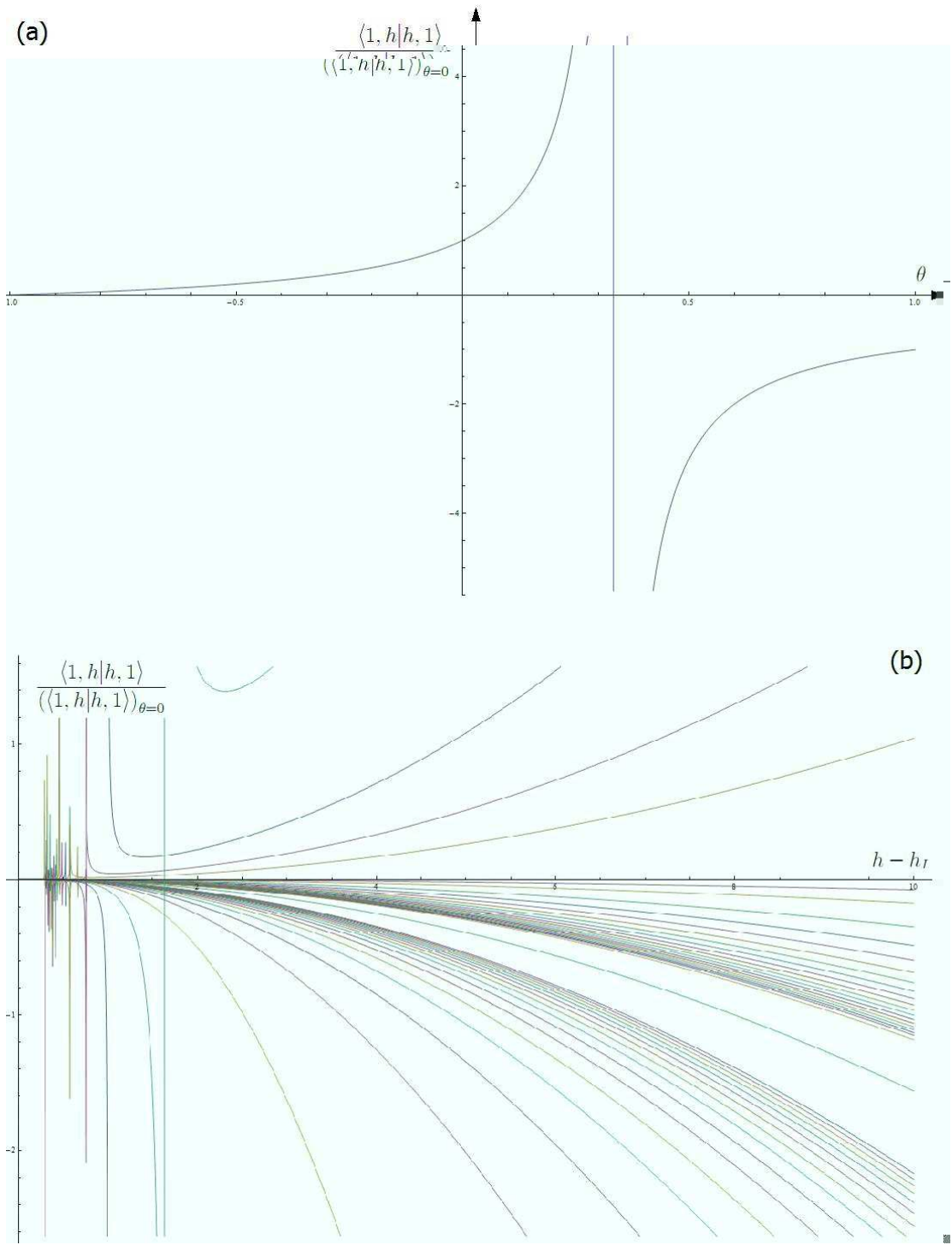}\\
  \caption{Graphics of (a) the normalized relative correlator (\ref{corr2}) (b) the correlator (\ref{corr3}) (vertical axis) as a function of the argument $h-h_I$ (horizontal axis) for some values of $\theta\in(-1,1)$.}\label{fig:cor}
\end{figure}

\begin{figure}
  \centering
  \includegraphics[scale=0.55]{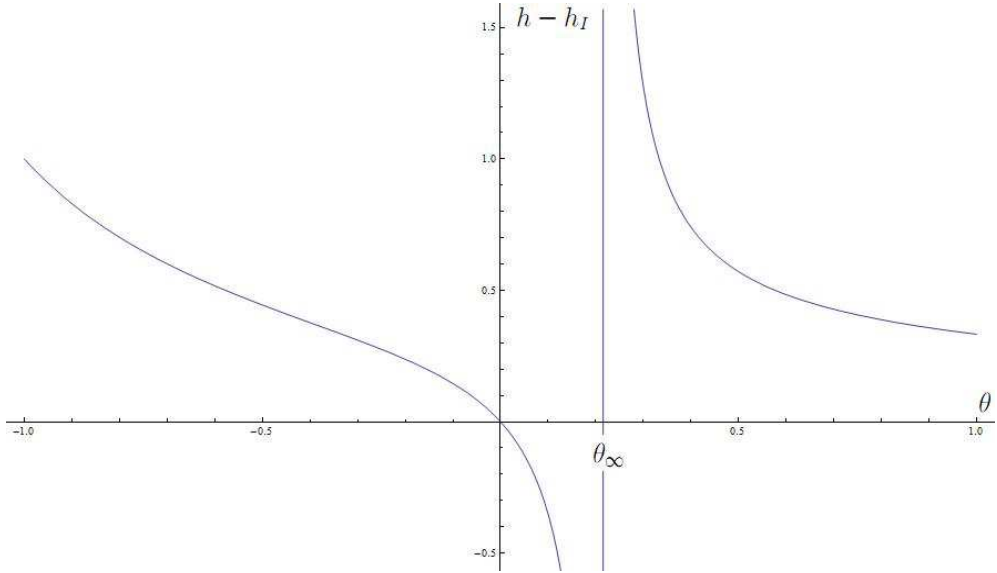}\\
 \caption{Dependence of the argument $h-h_I$ from the parameter $\theta$ for that the graviton is localized.}\label{fig:grav}
\end{figure}

Let us consider in detail the one point correlator (\ref{cor2}). Firstly let us note that by application of the correlator (\ref{cor2}) the bosonic field (\ref{field}) is immediately determined by this one-point correlator in the following way
\begin{eqnarray}\label{field1}
  \mathbf{\Psi}[h]=\frac{1}{2\sqrt{2m_I}}\sqrt{\langle 1,h|h,1\rangle}\left(e^{-i\theta[h]}\mathsf{G}_I+e^{i\theta[h]}\mathsf{G}^{\dagger}_I\right),
\end{eqnarray}
and by this it gives also the self-interaction
\begin{equation}
  \mathbf{\Psi}^\dagger[h]\mathbf{\Psi}[h]=\dfrac{\langle 1,h|h,1\rangle}{{_I}\langle 0|0\rangle_I}\left[\left(e^{-2i\theta[h]}\mathsf{G}_I+\mathsf{G}_I^\dagger\right)\mathsf{G}_I+h.c.\right],
\end{equation}
where $h.c.$ means hermitean conjugation. From the other side by using of the square of mass definition (\ref{masqr}) one can obtain the relation
\begin{equation}
\dfrac{2}{3h}\left(R[h]-2\Lambda-6\varrho\right)=m^2[h]=\dfrac{m_I^2}{\langle 1,h|h,1\rangle}=\dfrac{1}{64}\dfrac{{_I}\langle 0|0\rangle_I^2}{\langle 1,h|h,1\rangle},
\end{equation}
that can be interpreted as the other definition of the one-point correlator by basic geometrical quantities associated with 3-dimensional space metrics
\begin{equation}
  \dfrac{\langle 1,h|h,1\rangle}{{_I}\langle 0|0\rangle_I}=\dfrac{3}{128}\dfrac{h}{R[h]-2\Lambda-6\varrho}{_I}\langle 0|0\rangle_I.
\end{equation}
Also the Dark Matter energy density (\ref{dm}) can be expressed immediately by this correlator
\begin{equation}
  \rho_{DM}=\dfrac{3h}{788}\dfrac{{_I}\langle 0|0\rangle_I^2}{\langle 1,h|h,1\rangle}.
\end{equation}
Moreover, as it was noted in the previous subsection the monodromy matrix (\ref{mon}) is immediately determined by the quotient of squares of mass, that is really the one-point correlator by the formula (\ref{cor2}). In this manner the monodromy matrix (\ref{mon}) is really dependent on the one-point correlator in the following way
\begin{eqnarray}\label{mon1}
\mathbb{G}[h]\!=\!\left[\!\!\!\begin{array}{cc}
\dfrac{\sqrt{\langle 1,h|h,1\rangle}+1}{\sqrt[4]{\langle 1,h|h,1\rangle}}\dfrac{e^{-i\theta[h]}}{2}\vspace*{10pt}\!\!\!\!&
\dfrac{\sqrt{\langle 1,h|h,1\rangle}-1}{\sqrt[4]{\langle 1,h|h,1\rangle}}\dfrac{e^{i\theta[h]}}{2}\\
\dfrac{\sqrt{\langle 1,h|h,1\rangle}-1}{\sqrt[4]{\langle 1,h|h,1\rangle}}\dfrac{e^{-i\theta[h]}}{2}\!\!\!\!&
\dfrac{\sqrt{\langle 1,h|h,1\rangle}+1}{\sqrt[4]{\langle 1,h|h,1\rangle}}\dfrac{e^{i\theta[h]}}{2}\end{array}\!\!\!\right]\!\!,
\end{eqnarray}
where according to the definition (\ref{phas}) and with accepted sign, the phase is equal to
\begin{equation}
\theta[h]=m_I\int_{h_I}^{h}\sqrt{\langle 1,h|h,1\rangle}\delta h.
\end{equation}
Similarly the second quantization matrix (\ref{sqxm}) can be completely determined by the considered one-point correlator in the following way
\begin{equation}\label{sqxm1}
  \mathbb{Q}[h]=\left[\begin{array}{cc}\dfrac{1}{\sqrt{2m_I}}\sqrt[4]{\langle 1,h|h,1\rangle}&\dfrac{1}{\sqrt{2m_I}}\sqrt[4]{\langle 1,h|h,1\rangle}\\
-i\sqrt{\dfrac{m_I}{2}}\dfrac{1}{\sqrt[4]{\langle 1,h|h,1\rangle}}&i\sqrt{\dfrac{m_I}{2}}\dfrac{1}{\sqrt[4]{\langle 1,h|h,1\rangle}}\end{array}\right].
\end{equation}
Note that in the previous section we obtained the relation for square of mass (\ref{full}) by the coefficients $\alpha$'s
\begin{eqnarray}\label{full1} m^2[h]&=&-\dfrac{1}{(h-h_I)^2}+\left(\dfrac{\dfrac{\alpha_2}{h-h_I}}{1+\dfrac{\alpha_2}{h-h_I}}\right)^2\left[\dfrac{1}{(h-h_I)^2}-\dfrac{\alpha_1^2+2\alpha_1}{\alpha_2}\right]+\nonumber\\
&+&2\dfrac{\dfrac{\alpha_2}{h-h_I}}{\left(1+\dfrac{\alpha_2}{h-h_I}\right)^2}\left[\dfrac{1}{(h-h_I)^2}+\dfrac{\alpha_0}{\alpha_2}+\dfrac{\alpha_0}{h-h_I}\right],
\end{eqnarray}
and by this the correlator (\ref{cor2}) can be determined now by the following way
\begin{eqnarray}
  \dfrac{\langle 1,h|h,1\rangle}{{_I}\langle0|0\rangle_I^2}\!\!\!&=&\!\!\!\dfrac{1}{64}\Biggr\{-\dfrac{1}{(h-h_I)^2}+\left(\dfrac{\dfrac{\alpha_2}{h-h_I}}{1+\dfrac{\alpha_2}{h-h_I}}\right)^2\left[\dfrac{1}{(h-h_I)^2}-\dfrac{\alpha_1^2+2\alpha_1}{\alpha_2}\right]+\nonumber\\
\!\!\!&+&\!\!\!2\dfrac{\dfrac{\alpha_2}{h-h_I}}{\left(1+\dfrac{\alpha_2}{h-h_I}\right)^2}\left[\dfrac{1}{(h-h_I)^2}+\dfrac{\alpha_0}{\alpha_2}+\dfrac{\alpha_0}{h-h_I}\right]\Biggr\}^{-1},
\end{eqnarray}
that for case of constant energies of the rang $\varepsilon$ according to (\ref{add}) becomes
\begin{equation}\label{for}
\dfrac{\langle 1,h|h,1\rangle}{{_I}\langle0|0\rangle_I^2}=\dfrac{(h-h_I)^2}{64}\dfrac{\varepsilon^2(h-h_I)^3+4\varepsilon(h-h_I)^2+4(h-h_I)}{3\varepsilon^2(h-h_I)^3+8\varepsilon(h-h_I)^2-4(h-h_I)-4\varepsilon},
\end{equation}
and in the tachyon limit we obtain
\begin{equation}
\lim_{\varepsilon\rightarrow0}\dfrac{\langle 1,h|h,1\rangle}{{_I}\langle0|0\rangle_I^2}=-\dfrac{1}{64}(h-h_I)^2.
\end{equation}
The one-point correlation function (\ref{for}) can be interpreted as the information source on quantum stable states of the considered Bose system. Namely, as it is common accepted in research on similar situation in particle physics, one can consider the poles of the correlator with respect to the variable $h-h_I$. The poles are given by zeros of the correlator (\ref{for}) denominator
\begin{equation}
3\varepsilon^2(h-h_I)^3+8\varepsilon(h-h_I)^2-4(h-h_I)-4\varepsilon=0.
\end{equation}
This is the third order polynomial equation, and generally this kind of equations has three roots -- two complex and one real. Let us consider only the real root, because it is the stable state only. By elementary algebraic methods of the Galois group, one can obtain that the real solution of the equation (\ref{for}) is given by
\begin{eqnarray}
h-h_I&=&-\dfrac{8}{9\varepsilon}+\dfrac{50}{9\varepsilon\sqrt[3]{\dfrac{243}{4}\varepsilon^2-118+9\sqrt{3}\sqrt{-7-59\varepsilon^2+\dfrac{243}{16}\varepsilon^4}}}+\nonumber\\
&+&\dfrac{\sqrt[3]{-944+486\varepsilon^2+18\sqrt{3}\sqrt{-112-944\varepsilon^2+243\varepsilon^4}}}{9\varepsilon^2}.
\end{eqnarray}
This solution has a little bit complicated form, but we can exchange these some complex solution by the following recept. Namely, we know that the real solution is only one, and by this one can use the parametrization of the constant $\varepsilon$, related with $h-h_I$ by the inequality (\ref{ineq}), by the following way
\begin{equation}\label{reg}
  \varepsilon=\theta\dfrac{2}{h-h_I},
\end{equation}
where $\theta$ is a number, that by the formula (\ref{ineq}) must fulfills
\begin{equation}
  |\theta|<1\Rightarrow-1<\theta<1.
\end{equation}
With this supposition the formula (\ref{for}) can be expressed as follows
\begin{equation}\label{corr}
  \dfrac{\langle 1,h|h,1\rangle}{{_I}\langle0|0\rangle_I^2}=\left(\dfrac{\langle 1,h|h,1\rangle}{{_I}\langle0|0\rangle_I^2}\right)_{\theta=0}\dfrac{(\theta+1)^2(h-h_I)^2}{2\theta-(3\theta^2+4\theta-1)(h-h_I)^2},
\end{equation}
where the correlator
\begin{equation}
  \left(\dfrac{\langle 1,h|h,1\rangle}{{_I}\langle0|0\rangle_I^2}\right)_{\theta=0}=-\dfrac{(h-h_I)^2}{64},
\end{equation}
is identified with tachyon state. The real and positive pole of (\ref{corr}) is determined by very simple relation
\begin{equation}
h-h_I=\dfrac{2\theta}{3\theta^2+4\theta-1},
\end{equation}
In this manner, by using of the regularization (\ref{reg}), actually one can consider the relative correlator
\begin{equation}\label{corr3}
  \dfrac{\langle 1,h|h,1\rangle}{\left(\langle 1,h|h,1\rangle\right)_{\theta=0}}=\dfrac{(\theta+1)^2(h-h_I)^2}{2\theta-(3\theta^2+4\theta-1)(h-h_I)^2},
\end{equation}
where the vacuum-vacuum amplitude was absorbed by reduction. This correlator expressed in units for which the square of the tachyon mass is equal to minus unity, \emph{i.e.} for
\begin{equation}
  m_0^2=-\dfrac{1}{(h-h_I)^2}\equiv-1,
\end{equation}
becomes very simple
\begin{equation}\label{corr2}
  \left(\dfrac{\langle 1,h|h,1\rangle}{\left(\langle 1,h|h,1\rangle\right)_{\theta=0}}\right)_{m_0^2=-1}=\dfrac{1+\theta}{1-3\theta},
\end{equation}
and we see that this reduced one-point correlator has the real pole for $\theta=\dfrac{1}{3}$. The Figure (\ref{fig:cor}) presents graphics of this one-point correlator as a function of $\theta$, and the one-point correlator (\ref{corr3}) as function of the argument $h-h_I$ for some values of $\theta\in(-1,1)$.
Because we have to deal with one-point correlation function of the quantum bosonic field $\mathbf{\Psi}$ determined on the configurational space of General Relativity that is the superspace, where the point means some concrete compact 3-geometry, the poles of the one-point correlator (\ref{corr2}) have an interpretation of free stable states of the considered quantum field theory. By this the point $\theta=\dfrac{1}{3}$ localizes the quantum stable state, and in this manner the quantum stable state can be identified with the quanta of gravity, \emph{i.e.} with the graviton. Generally the graviton is detected in the superspace points $h$ that fulfill the relation
\begin{equation}\label{pol}
h-h_I=\dfrac{2\theta}{3\theta^2+4\theta-1},
\end{equation}
and are presented on the Figure (\ref{fig:cor}) in the part (b) by points where the correlator has singularity. In the tachyon limit $\theta=0$ the pole value of $h-h_I$ (\ref{pol}) becomes
\begin{equation}
(h-h_I)_{\theta=0}=0,
\end{equation}
and by this in the point $h=h_I$ is localized graviton in the tachyon limit. It is interesting that in the point
\begin{equation}
  \theta=\theta_{\infty}=\dfrac{\sqrt{7}-2}{3}\approx0.2152504370,
\end{equation}
stable quantum state can not be detected, as it is presented on the Figure (\ref{fig:grav}).
\newpage
\section{Classical bosonic embedded space. Tachyon.}
Let us consider now some aspects of the classical Bose field $\Psi[h]$ introduced in the previous section. We are not going to resolve the equation (\ref{new}) again, but analyze some structural elements of the relativistic quantum mechanics described by this equation, especially the square of mass (\ref{masqr}) by its direct connection with 3-geometry of spacetime.
\subsection{Field mass by its energy}
The relativistic wave equation obtained in the previous section
\begin{equation}\label{kgf}
  \left\{\dfrac{\delta^2}{\delta h^2}+m^2[h]\right\}\Psi[h]=0,
\end{equation}
from classical point of view describes some one-dimensional classical particle with mass dependent on the point $h_{ij}$ in superspace characterized by its determinant $h$. In the case of the mass independent on the superspace point, this equation has a very well known solution in terms of plane waves, but in the general case, \emph{i.e.} for nonconstant mass, this equation is not very simple for direct solving. Plane waves are not a solution in this case. However, we are not going to concentrate our considerations on search for general classical solutions of the equation (\ref{kgf}), but in the next sections we will try to construct a second quantization of this equation, that is independent on classical field theory solution. In this section we will discuss classical field theory that gives the equation (\ref{kgf}) as the classical Euler--Lagrange equation of motion. Obviously, because of we have to deal with the Bose field, in this section we will consider some theory of the bosonic string.

Firstly, let us consider the equation (\ref{kgf}) as the classical Euler--Lagrange equation of motion obtained by some field theory lagrangian $L\left[\Psi[h],\dfrac{\delta\Psi[h]}{\delta h}\right]$ according to the system of equations
\begin{eqnarray}
  \dfrac{\delta \Pi_{\Psi}[h]}{\delta h}-\dfrac{\partial}{\partial \Psi[h]}L\left[\Psi[h],\dfrac{\delta\Psi[h]}{\delta h}\right]&=&0,\\
  \Pi_{\Psi}[h]-\dfrac{\partial}{\partial \left(\dfrac{\delta\Psi[h]}{\delta h}\right)} L\left[\Psi[h],\dfrac{\delta\Psi[h]}{\delta h}\right]&=&0,
\end{eqnarray}
where $\Pi_\Psi[h]$ is canonical momentum conjugate to the classical field $\Psi[h]$. Standardly, one can construct the Lagrangian by field theory action functional $S[\Psi]$ directly by using of the equation of motion (\ref{kgf}) in the following way
\begin{eqnarray}\label{lag}
  S[\Psi]&=&-\dfrac{1}{2}\int\delta{h}\Psi[h]\left\{\dfrac{\delta^2}{\delta{h}^2}+m^2[h]\right\}\Psi[h]\nonumber=\\
&=&-\dfrac{1}{2}\int\delta{h}\left\{\dfrac{\delta}{\delta{h}}\left(\Psi[h]\dfrac{\delta\Psi[h]}{\delta{h}}\right)-\left(\dfrac{\delta\Psi[h]}{\delta{h}}\right)^2+m^2[h]\Psi^2[h]\right\}=\nonumber\\
&=&\dfrac{1}{2}\int\delta{h}\left\{\left(\dfrac{\delta\Psi[h]}{\delta{h}}\right)^2-m^2[h]\Psi^2[h]\right\}\equiv\int\delta h L\left[\Psi[h],\dfrac{\delta\Psi[h]}{\delta h}\right]\!\!,
\end{eqnarray}
where we have applied integration of full divergence, and $\Psi^\dagger[h]=\Psi[h]$. This means we have to deal with the classical field theory given by the Euler--Lagrange system of equations in the form
\begin{eqnarray}
\dfrac{\delta \Pi_{\Psi}[h]}{\delta h}+m^2[h]\Psi[h]&=&0,\label{lagr1}\\
\Pi_{\Psi}[h]-\dfrac{\delta\Psi[h]}{\delta h}&=&0.\label{lagr2}
\end{eqnarray}
The classical field theory Hamiltonian can be constructed immediately from the Lagrangian (\ref{lag}) by application of the standard Legendre transformation \cite{gol} between Hamiltonian and Lagrangian of the classical dynamical system
\begin{equation}\label{ham}
  H[\Pi_\Psi,\Psi]=\Pi_{\Psi}[h]\dfrac{\delta\Psi[h]}{\delta h}-L\left[\Psi[h],\dfrac{\delta\Psi[h]}{\delta h}\right]=\dfrac{\Pi_\Psi^2[h]+m^2[h]\Psi^2[h]}{2},
\end{equation}
that for fixed mass $m[h]$ describes ellipse in space $(H[\Pi_\Psi,\Psi],\Pi_\Psi,\Pi)$. The set of these all ellipses lies on paraboloid parameterized by continue parameter $h$. Direct using of the momentum constraint (\ref{lagr2}) and the fact that really the Hamiltonian (\ref{ham}) can be treated as functional with respect to the evolution parameter $h$ of the equation (\ref{kgf}), \emph{i.e.} $H[\Pi_\Psi[h],\Psi[h]]=H[h]$, allows to rewrite the classical field theory Hamiltonian  (\ref{ham}) as the definition of mass
\begin{equation}\label{m}
m^2[h]\Psi^2[h]=2H[h]-\left(\dfrac{\delta\Psi[h]}{\delta h}\right)^2,
\end{equation}
and after simple elimination of the square of mass by using of the equation of motion (\ref{kgf}) this leads to the functional differential equation for the classical field $\Psi[h]$
\begin{equation}\label{r1}
\dfrac{\delta^2\Psi[h]}{\delta h^2}\Psi[h]=2H[h]-\left(\dfrac{\delta\Psi[h]}{\delta h}\right)^2,
\end{equation}
which after collecting terms leads to the following relation between the field $\Psi[h]$ and the Hamiltonian $H[h]$
\begin{equation}
  \dfrac{\delta}{\delta h}\left(\dfrac{\delta\Psi[h]}{\delta h}\Psi[h]\right)=2H[h].
\end{equation}
So, presently one can be integrate the last equation directly with initial value of $h$ taken as $h_I$. In result we obtain
\begin{equation}\label{int}
\Psi^2[h]=4\int_{h_I}^h\delta h'\int_{h_I}^{h'}\delta h''H[h''],
\end{equation}
and by this the solution of the classical wave equation (\ref{kgf}) can be formally accepted as the functional $\Psi[h]=\left(\Psi^2[h]\right)^{1/2}$. From the other side the equation (\ref{r1}) can be integrated into the form
\begin{equation}\label{r2}
\Pi_{\Psi}[h]\Psi[h]=2\int_{h_I}^h\delta h'H[h'],
\end{equation}
which combined together with the solution (\ref{int}) fixes values of the canonical conjugate momentum $\Pi_\Psi[h]$ with respect to values of the classical Hamiltonian $H[h]$ in the following way
\begin{equation}
\Pi_\Psi[h]=\dfrac{\int_{h_I}^h\delta h'H[h']}{\left(\int_{h_I}^h\delta h'\int_{h_I}^{h'}\delta h''H[h'']\right)^{1/2}}.
\end{equation}
Taking into account the basic relation for the classical field theory Hamiltonian (\ref{ham}) one can obtain by direct algebraic manipulation
\begin{equation}
  m^2[h]=\dfrac{2H[h]}{\Psi^2[h]}-\left(\dfrac{\Pi_{\Psi}[h]}{\Psi[h]}\right)^2,
\end{equation}
or by employing the relation (\ref{r2})
\begin{equation}
  m^2[h]=\dfrac{2H[h]}{\Psi^2[h]}-\left(\dfrac{2\int_{h_I}^h\delta h'H[h']}{\Psi^2[h]}\right)^2.
\end{equation}
By treating this relation as a constraint that fixes mass value and by application of the solution (\ref{int}) one can determine easily the dependence between field mass and its energy values
\begin{equation}\label{mascon}
  m^2[h]=\dfrac{1}{4}\left[\dfrac{2H[h]}{\int_{h_I}^h\delta h'\int_{h_I}^{h'}\delta h''H[h'']}-\left(\dfrac{\int_{h_I}^h\delta h'H[h']}{\int_{h_I}^h\delta h'\int_{h_I}^{h'}\delta h''H[h'']}\right)^2\right].
\end{equation}
In this manner, the classical field theory of the Bose field $\Psi[h]$ can be studied in terms of values of its square mass in dependence on values of the classical field theory Hamiltonian.
\newpage
\subsection{Perturbations, cosmological constant, Dark Energy}
The last formula (\ref{mascon}) determines fundamental relation between the square of mass of the considered boson and the classical field energy distribution. By using of the definition (\ref{masqr}) one can consider this relation in terms of the constraint
\begin{equation}
\dfrac{8}{3h}\left(R[h]-2\Lambda-6\varrho\right)=\dfrac{2H[h]}{\int_{h_I}^h\delta h'\int_{h_I}^{h'}\delta h''H[h'']}-\left(\dfrac{\int_{h_I}^h\delta h'H[h']}{\int_{h_I}^h\delta h'\int_{h_I}^{h'}\delta h''H[h'']}\right)^2,
\end{equation}
that fixes values of the normal to the boundary space component of the stress--energy tensor $\varrho$ on
\begin{eqnarray}
  \varrho\!\!&=&\!\!\dfrac{h}{16}\left[\left(\dfrac{\int_{h_I}^h\delta h'H[h']}{\int_{h_I}^h\delta h'\int_{h_I}^{h'}\delta h''H[h'']}\right)^2-\dfrac{2H[h]}{\int_{h_I}^h\delta h'\int_{h_I}^{h'}\delta h''H[h'']}\right]+\nonumber\\
\!\!&+&\!\!\dfrac{R[h]}{6}-\dfrac{\Lambda}{3},
\end{eqnarray}
and by positive definiteness of the classical energy density can be used to determine an upper limit for the cosmological constant
\begin{equation}
  \Lambda\leq\dfrac{R[h]}{2}+\dfrac{3h}{8}\left[\left(\dfrac{\int_{h_I}^h\delta h'H[h']}{\int_{h_I}^h\delta h'\int_{h_I}^{h'}\delta h''H[h'']}\right)^2-\dfrac{2H[h]}{\int_{h_I}^h\delta h'\int_{h_I}^{h'}\delta h''H[h'']}\right]
\end{equation}

Let us define the mass groundstate of the classical field theory presented above by the following condition
\begin{equation}
 H[h']=C\delta(h'-h),~~H[h]=0,
\end{equation}
so that the constant $C\rightarrow0$ formally, here $\delta(h'-h)$ is the Dirac delta function. For so defined groundstate the first term in the square of mass (\ref{mascon}) vanishes automatically, and the second term gives finite contribution to the square of mass
\begin{equation}\label{ground}
m^2_0[h]\equiv m^2[h]\Big|_{\mathrm{groundstate}}=-\left(\dfrac{C}{C(h-h_I)}\right)^2\Bigg|_{C\rightarrow0}=-\dfrac{1}{4\left(h-h_I\right)^2}.
\end{equation}
This number is negative for all values of $h$ and by this relation describes formally the tachyon, that is the fundamental excitation of the bosonic string \cite{str}. For the considered groundstate are fulfilled the following relations
\begin{eqnarray}
  \varrho^{(0)}&=&\dfrac{1}{16}\dfrac{h}{\left(h-h_I\right)^2}+\dfrac{R[h]}{6}-\dfrac{\Lambda}{3},\\
\Lambda&\leq&\dfrac{R[h]}{2}+\dfrac{1}{16}\dfrac{h}{\left(h-h_I\right)^2}.
\end{eqnarray}
If we demand additionally that for the initial metric $h_I$ the classical boson field $\Psi[h]$ should have a some finite mass $m_I$ then the formula (\ref{tach}) should be renormalized as follows
\begin{equation}\label{tach1}
  m^2_0[h]=-\dfrac{1}{\left(h-h_I-i\sqrt{\dfrac{1}{m_I^2}}\right)^2},
\end{equation}
so the initial square of mass should be huge for correctness $\eta=\sqrt{\dfrac{1}{m_I^2}}\rightarrow0$.
For the field $\Psi[h]$, we conclude from the basic relation (\ref{int}) that in so defined mass groundstate the classical field $\Psi[h]$ is
\begin{equation}
  \Psi[h]=2\sqrt{C(h-h_I)},
\end{equation}
and in this case the phase space $(\Pi_\Psi,\Psi)$ determined by the relation (\ref{r2}) is given by a family of hyperbolas
\begin{equation}
  \Pi_\Psi[h]=\dfrac{2C}{\Psi[h]}\equiv\sqrt{\dfrac{C}{h-h_I}},
\end{equation}
or simply by the condition that the product of phase space variables is the first integral of the considered classical field theory $\Pi_\Psi[h]\Psi[h]=constans$.

For all constant, but nonzero values of the classical field theory Hamiltonian $H[h]=H_0\neq0$, the square of mass vanishes identically
\begin{equation}\label{tach}
  m^2[h]\Big|_{H[h]=H_0}=0,
\end{equation}
and these states are massless excitations of the bosonic string, by fact that here $H_0$ is arbitrary constant, number of massless states is continuum. For the massless states we have simplified relations for normal stress--energy tensor and cosmological constant
\begin{equation}
  \varrho=\dfrac{R[h]}{6}-\dfrac{\Lambda}{3},~~\Lambda\leq\dfrac{R[h]}{2}.
\end{equation}
However, presence of massless states in the theory means that $\mu^2=0$, what is unphysical mass scale value by $\mu\geq1$.

From the string theory point of view the tachyon state is treated as mass groundstate of the considered theory of bosonic string. One can generate the process of symmetry breaking in frames of the perturbational calculus with respect to the classical field theory Hamiltonian $H[h]$. Namely, in the most general case, one can imagine that an arbitrary mass state of the considered bosonic string, and as the context suggests arbitrary metrics of General Relativity, is generated by small deviation from the tachyon state. Let the deviation is an arbitrary functional so that $\delta H[h]\ll1$, then deviation from the groundstate of the classical Hamiltonian given by
\begin{equation}\label{ssb}
H[h']=C\delta(h'-h)+\delta H[h'],~~H[h]=\delta H[h],
\end{equation}
leads to perturbations from the mass groundstate in the form
\begin{equation}\label{ssb1}
m^2[h]=m^2_{0}[h]+\delta m^2[h],
\end{equation}
where the term $\delta m^2[h]$ describes the full contribution to the square of mass from the perturbation and breaks mass groundstate directly. Let us assume that the term has a form of the series
\begin{equation}\label{ssb2}
  \delta m^2[h]=\sum_{n=1}^{\infty}\delta^{(n)}m^2[h]=\delta^{(1)}m^2[h]+\delta^{(2)}m^2[h]+\delta^{(3)}m^2[h]+\ldots,
\end{equation}
where the partial terms $\delta^{(k)}m^2[h]$ consist all corrections taken up to the $k$-th order in the perturbation $\delta H[h]$ of the classical Hamiltonian. By introduce of shorten notation
\begin{eqnarray}
\alpha_0\equiv \alpha_0[h]&=&\dfrac{\delta H[h]}{C},\label{d0}\\
\alpha_1\equiv \alpha_1[h]&=&\int_{h_I}^h\delta h'\alpha_0[h'],\label{d1}\\
\alpha_2\equiv \alpha_2[h]&=&\int_{h_I}^{h}\delta h'\alpha_1[h']=\int_{h_I}^h\delta h'\int_{h_I}^{h'}\delta h''\alpha_0[h''],\label{d2}
\end{eqnarray}
one can check easily by elementary computation that the $k$-th contribution to the series (\ref{ssb2}) has a following form
\begin{eqnarray}\label{cormas}
\delta^{(k)} m^2[h]&=&(-1)^{k+1}\dfrac{k+1}{4(h-h_I)^{k+2}}\alpha_2^k+(-1)^k\dfrac{2k}{4(h-h_I)^{k+1}}\alpha_1\alpha_2^{k-1}+\nonumber\\
&+&(-1)^{k-1}\dfrac{2\alpha_0\alpha_2^{k-1}+(k-1)\alpha_1^2\alpha_2^{k-2}}{4(h-h_I)^k},
\end{eqnarray}
and by this the series (\ref{ssb2}) can be summed immediately, so that the full result for the square of mass (\ref{ssb1}) can be determined by dependence from the parameters $\alpha's$
\begin{eqnarray}\label{full} m^2[h]&=&-\dfrac{1}{4(h-h_I)^2}+\left(\dfrac{\dfrac{1}{2}\dfrac{\alpha_2}{h-h_I}}{1+\dfrac{\alpha_2}{h-h_I}}\right)^2\left[\dfrac{1}{(h-h_I)^2}-\dfrac{\alpha_1^2+2\alpha_1}{\alpha_2}\right]+\nonumber\\
&+&\dfrac{\dfrac{1}{2}\dfrac{\alpha_2}{h-h_I}}{\left(1+\dfrac{\alpha_2}{h-h_I}\right)^2}\left[\dfrac{1}{(h-h_I)^2}+\dfrac{\alpha_0}{\alpha_2}+\dfrac{\alpha_0}{h-h_I}\right],
\end{eqnarray}
and the parameters can be treated as free parameters of the theory.
\subsection{Tachyon}
\begin{figure}
\centering
  \includegraphics[scale=0.75]{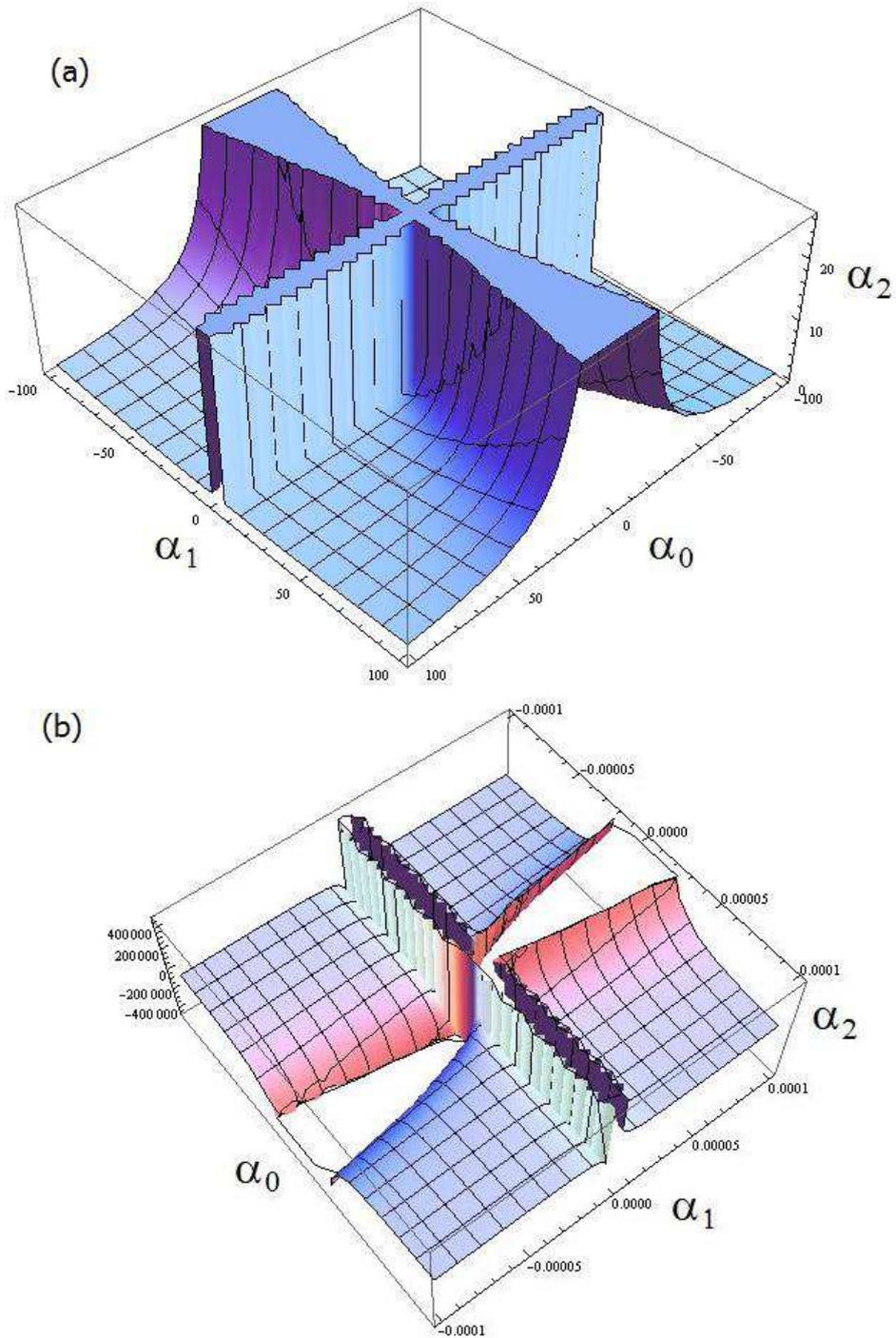}\\
  \caption{Tachyon state in space of parameters $(\alpha_0,\alpha_1,\alpha_2)$: the part (a) presents large scale view of the surface (\ref{tachyon}); the part (b) presents the surface in neighborhood of the point $(0,0,0)$. }\label{fig:tachyon}
\end{figure}

From the relation (\ref{full}) we see explicitly that if $\alpha$'s are constrained by the following system of equations
\begin{equation}\label{soe}
  \left\{\begin{array}{cc}\dfrac{1}{(h-h_I)^2}-\dfrac{\alpha_1^2+2\alpha_1}{\alpha_2}&=0\\
  \dfrac{1}{(h-h_I)^2}+\dfrac{\alpha_0}{\alpha_2}+\dfrac{\alpha_0}{h-h_I}&=0\end{array}\right.
\end{equation}
then we have to deal with the tachyon state -- in this case the square of mass is negative and equal to the first term of this formula. The system of equations (\ref{soe}) can be solved directly, in result we obtain the relations between $\alpha$'s
\begin{eqnarray}
  \alpha_1&=&\pm\sqrt{1-\dfrac{\alpha_0}{1+\alpha_0(h-h_I)}}-1,\label{a1}\\
\alpha_2&=&-\dfrac{\alpha_0(h-h_I)^2}{1+\alpha_0(h-h_I)}~~\mathrm{for}~~\mathrm{both}~~\alpha_1.\label{a2}
\end{eqnarray}
By this the tachyon, which is the mass groundstate of the considered bosonic theory, can be completely determined by arbitrary value of $\alpha_0$ and connected with this value the functions $\alpha_1$ and $\alpha_2$ determined by relations (\ref{a1}) and (\ref{a2}). One can see easily that this system of equations leads to the surface $T$ in space of parameters $(\alpha_0,\alpha_1,\alpha_2)$ given by the set of points
\begin{equation}\label{tachyon}
T=\left\{(\alpha_0,\alpha_1,\alpha_2)\in{\mathbb{R}^3}:\alpha_2(\alpha_0,\alpha_1)=\dfrac{\left(\alpha_0+2\alpha_1+\alpha_1^2\right)^2}{\alpha_0^2\left(2\alpha_1+\alpha_1^2\right)}\right\},
\end{equation}
that describes tachyon state in this space, see Figure \ref{fig:tachyon}.

Consider the case of the constant perturbation $\epsilon$ that is very small in comparison with $C$
\begin{equation}
  \dfrac{\int_{h_I}^{h}\delta h'\delta H[h']}{\int_{h_I}^{h}\delta h'H[h']}=\dfrac{\epsilon}{C}=\varepsilon\ll 1.
\end{equation}
Then by direct combination of the relations (\ref{d0}), (\ref{d1}), and (\ref{d2}) we obtain
\begin{eqnarray}
  \alpha_0&=&\varepsilon,\\
  \alpha_1&=&(h-h_I)\varepsilon,\\
  \alpha_2&=&\dfrac{(h-h_I)^2}{2}\varepsilon=\dfrac{\alpha_1^2}{2\alpha_0},
\end{eqnarray}
and in this case
\begin{eqnarray}
  \delta^{(k)}m^2[h]=(-1)^{k-1}\dfrac{(h-h_I)^{k-2}}{2^{k}}\left[2k+4-\dfrac{2k}{(h-h_I)^2}\right]\varepsilon^k,
\end{eqnarray}
so, the sum (\ref{ssb1}) can be calculated directly
\begin{equation}\label{add}
  m^2[h]=-\dfrac{1}{(h-h_I)^2}+\dfrac{4\varepsilon}{(h-h_I)^3}\dfrac{\varepsilon(h-h_I)^3+3(h-h_I)^2-1}{\varepsilon^2(h-h_I)^2+4\varepsilon(h-h_I)+4},
\end{equation}
where the small constant $\varepsilon$ is chosen according to the condition
\begin{equation}\label{ineq}
  |\varepsilon|<\dfrac{2}{|h-h_I|}.
\end{equation}
It is clear now that tachyon state is obtained by the limit $\varepsilon\rightarrow0$.

Equivalently one can treat the square of mass (\ref{ssb1}) in terms of power series in the function $\dfrac{1}{h-h_I}$
\begin{equation}\label{ser}
m^2[h]=\sum_{n=0}^{\infty}\dfrac{a_n[h;h_I]}{\left(h-h_I\right)^n},
\end{equation}
where the coefficients $a_n$ as functions of (\ref{d0}), (\ref{d1}), and (\ref{d2}) are functionals of $h$ and initial data described by $h_I$, and they can be directly written in the compact form
\begin{eqnarray}\label{coeff}
a_n[h;h_I]=(-1)^{n}\left\{\left[2\alpha_0\alpha_2-\left(1+\alpha_1\right)^2\right]G[n]-2n\alpha_0\alpha_2\right\}\alpha_2^{n-2},
\end{eqnarray}
where $G[n]$ is the step function defined as
\begin{equation}
  G[n]=\left\{\begin{array}{cc}0,&\mathrm{for}~n<1\\n-1,&\mathrm{for}~n\geq1\end{array}\right.
\end{equation}

By using of the main relation for the square of mass (\ref{masqr})
\begin{equation}\nonumber
m^2[h]=\dfrac{2}{3h}\left(R[h]-2\Lambda-6\varrho\right),
\end{equation}
one can obtain by direct comparison with the power series (\ref{ser}) the relation
\begin{equation}\label{rel1}
\dfrac{2}{3h}\left(R[h]-2\Lambda-6\varrho\right)=\sum_{n=0}^{\infty}\dfrac{a_n[h;h_I]}{\left(h-h_I\right)^n},
\end{equation}
that can be treated as a definition of the stress--energy tensor $\varrho$ projected onto normal vector field to boundary 3-dimensional surface as
\begin{equation}\label{set}
\varrho[h]=\dfrac{R[h]}{6}-\dfrac{\Lambda}{3}-\dfrac{h}{4}\sum_{n=0}^{\infty}\dfrac{a_n[h;h_I]}{\left(h-h_I\right)^n}.
\end{equation}
This energy density is positive iff
\begin{equation}
  \Lambda\leq\dfrac{R[h]}{2}-\dfrac{3h}{4}\sum_{n=0}^{\infty}\dfrac{a_n[h;h_I]}{\left(h-h_I\right)^n},
\end{equation}
and this actually defines an upper limit for the cosmological constant $\Lambda$.

One can view on the relation (\ref{rel1}) by different point of view. Namely, when we rewrite this formula in the following form
\begin{equation}\label{dm1}
\dfrac{2}{3h}\left(R[h]-2\Lambda-6\varrho\right)-\sum_{n=0}^{\infty}\dfrac{a_n[h;h_I]}{\left(h-h_I\right)^n}=0,
\end{equation}
then we see that this suggests redefinition of the energy density by the way
\begin{equation}\label{tdm}
 T^{DM}=\varrho+\rho_{DM}[h],
\end{equation}
where
\begin{eqnarray}\label{dm}
  \rho_{DM}[h]&=&\dfrac{h}{4}\sum_{n=0}^{\infty}\dfrac{a_n[h;h_I]}{\left(h-h_I\right)^n},
\end{eqnarray}
can be interpreted as a density energy from Dark Matter fields. Equivalently one can determine the Dark Matter density energy (\ref{dm}) by application the relation (\ref{full}) as follows
\begin{eqnarray}
  \rho_{DM}[h]&=&\dfrac{h}{16}\Biggr\{-\dfrac{1}{(h-h_I)^2}+\left(\dfrac{\dfrac{\alpha_2}{h-h_I}}{1+\dfrac{\alpha_2}{h-h_I}}\right)^2\left[\dfrac{1}{(h-h_I)^2}-\dfrac{\alpha_1^2+2\alpha_1}{\alpha_2}\right]+\nonumber\\
&+&\dfrac{2\dfrac{\alpha_2}{h-h_I}}{\left(1+\dfrac{\alpha_2}{h-h_I}\right)^2}\left[\dfrac{1}{(h-h_I)^2}+\dfrac{\alpha_0}{\alpha_2}+\dfrac{\alpha_0}{h-h_I}\right]\Biggr\},
\end{eqnarray}
that for constant energies becomes
\begin{eqnarray}
  \rho_{DM}[h]=\dfrac{h}{4}\dfrac{\varepsilon}{(h-h_I)^3}\dfrac{\varepsilon(h-h_I)^3+3(h-h_I)^2-1}{\varepsilon^2(h-h_I)^2+4\varepsilon(h-h_I)+4}.\end{eqnarray}
By direct resolving of the equation (\ref{dm1}) with respect to the cosmological constant $\Lambda$ one can determine the cosmological constant as the quantity dependent only on scalar curvature of boundary 3-geometry, and summarized energy density of normal Matter fields and Dark Matter
\begin{equation}\label{lam}
  \Lambda=\dfrac{R[h]-3T^{DM}}{2}.
\end{equation}
In this manner, the Einstein--Hilbert action that is the second integral of (\ref{eh0}) takes the form
\begin{equation}
  S_{EH}[g]\longrightarrow S_{M+DM}[g]=\int_{M}d^4x\sqrt{-g}\left\{-\dfrac{1}{6}R[g]+\mathcal{L}_{M+DM}\right\},
\end{equation}
where $\mathcal{L}_{M+DM}$ is the total lagrangian of Matter fields and Dark Matter
\begin{equation}
  \mathcal{L}_{M+DM}=\mathcal{L}+\dfrac{R[h]}{6}-\dfrac{T^{DM}}{2},
\end{equation}
with $\mathcal{L}$ as the Lagrangian of Matter fields. Moreover, by constant value of $\Lambda$, with Dark Matter contribution, the General Relativity field equations (\ref{gr1}) presently are
\begin{equation}\label{grdm}
R_{\mu\nu}=\left[\dfrac{1}{2}R[h]-\dfrac{3}{2}\left(\varrho+\rho_{DM}[h]\right)\right]g_{\mu\nu}+3\left(T_{\mu\nu}-\dfrac{T}{2}g_{\mu\nu}\right),
\end{equation}
where $T_{\mu\nu}$ is the stress-energy tensor of Matter fields.

One can consider the case when we have to deal with vanishing cosmological constant $\Lambda\equiv0$. In this case, from the relations (\ref{lam}) and (\ref{tdm}) we directly obtain that stress-energy tensor projected onto normal field vector has a value
\begin{equation}
  \lim_{\Lambda\rightarrow0}\varrho[h]=\dfrac{R[h]}{3}-\rho_{DM}[h],
\end{equation}
that for small energies becomes
\begin{equation}
  \lim_{\Lambda\rightarrow0}\varrho[h]=\dfrac{R[h]}{3}-\dfrac{h}{4(h-h_I)^3}\dfrac{\varepsilon^2(h-h_I)^3+\varepsilon[3(h-h_I)^2-1]}{\varepsilon^2(h-h_I)^2+4\varepsilon(h-h_I)+4},
\end{equation}
and in the tachyon limit takes the value
\begin{equation}
  \lim_{\varepsilon\rightarrow0}\lim_{\Lambda\rightarrow0}\varrho[h]=\dfrac{R[h]}{3}.
\end{equation}
By this the Einstein--Hilbert field equations of General Relativity with Dark Matter existence (\ref{grdm}) in the case of vanishing cosmological constant within the tachyon limit are simply
\begin{equation}
  R_{\mu\nu}=3\left(T_{\mu\nu}-\dfrac{T}{2}g_{\mu\nu}\right),
\end{equation}
but the tachyon state in the neighborhood of zero in the space of parameters $(\alpha_0,\alpha_1,\alpha_2)$ with nonzero cosmological constant is described by completely other field equations
\begin{equation}
R_{\mu\nu}=\left(\dfrac{1}{2}R[h]-\dfrac{3}{2}\varrho\right)g_{\mu\nu}+3\left(T_{\mu\nu}-\dfrac{T}{2}g_{\mu\nu}\right).
\end{equation}
Perturbation calculus ideas presented above, completely describe the classical field theory (\ref{kgf}) in terms of the spontaneously breaking of mass groundstate of the bosonic string with respect to the field theory Hamiltonian (\ref{ham}).
\newpage
\section{Quantum gravity by thermodynamics}

The last section was devoted to presentation of the quantum theory of the Bose field $\mathbf{\Psi}[h]$ based on the quantization in the Fock space of creation and annihilation functional operators, and proper choice of the initial data basis. This approach led us to notion of space quantum states associated with three-dimensional spatial part of a Riemannian spacetime classically treated as a solution of the Einstein--Hilbert field equations of General Relativity. Furthermore, as the main result of our studies of the  one-point two-field correlator real poles of the bosonic field $\mathbf{\Psi}[h]$ we have obtained a localization of the stable quantum states of the quantum field theory that can be interpreted as the quantum particle of generalized gravitational fields -- the graviton.

In this section we will investigate thermodynamical description of the considered bosonic statistical system. We will use the density functional method in order to formulation of equilibrium statistical thermodynamics of many space quantum states. Actually, it is the last step of the Thermodynamical Einstein's Dream, that is the main motivation to this paper.

Let us try to create thermodynamical picture that arises from the presented quantum field theory. When we build thermodynamics, we should use the simplest rules of statistical physics, that in some sense give the general information about the considered physical system. In usual thermal situations in physics, we have to deal with some concrete set of possible physical states, and we try to construct statistical description of the system by using of ensemble that given the prescription for averaging procedure. Generally in real physical systems we have to deal with the only one classical statistics, \emph{i.e.} the Boltzmann distribution, and in case of quantum states of the Bose systems with the Bose--Einstein statistics, and in case with Fermi particles with Fermi--Dirac statistics. Furthermore, the real systems are no isolated and open, so interaction with environment is inevitable. Let us consider the situation of the concrete system as is the system of space quantum states of a spacetime. By quantum character of the set we have to deal with quantum statistics, in the considered case the quantum mechanics, that is classical field theory, is described by the Wheeler--DeWitt equation in form of the Klein--Gordon--Fock evolution equation (\ref{kgf}). Naturally, this is the Bose system, and we should describe statistical properties of the system in frames of the Bose--Einstein statistics. Moreover, by its Nature the system is open and no isolated, but we have proposed the diagonalization procedure and this framework generates the fundamental static operator basis in the Fock space associated with initial data of the system. Actually, this initial data basis also defines the thermal equilibrium state, and generalized thermodynamics of the set of space quantum states can be investigated from this point of view. Let us consider the thermodynamics of space quantum states as quantum theory of general gravitational fields.

The initial data basis (\ref{in}) gives an opportunity to introduce a notion of the thermodynamical equilibrium state in the statistical ensemble of many space quantum states that are some generalized quantum particles of the classical Einstein--Hilbert Riemannian manifold of General Relativity. Essentially, the fundamental static operator basis $\mathfrak{B}_I$ is given by creation and annihilation operators in the Fock space of the quantum field theory. It means that initial data are directly jointed with static description of the ensemble, and from the point of view of the fundamental basis the set of space quantum states is isolated and no open system, and can be characterized by usual thermodynamical description. By this from conceptual side of thermodynamics as the only theory between quantum field theory and statistical mechanics, and from as logical well as ontological points should be possible to obtain the statistical characterizations of the space quantum states system. In the context of this paper the following supposition seems to be the most natural
\begin{center}
\begin{tabular}{|p{10.5cm}|}
\hline \vspace*{0pt}\textsf{Thermodynamics of space quantum states is quantum gravity.}\vspace*{10pt}\\ \hline
\end{tabular}
\end{center}
Let us study this generalized thermodynamics and its physical aspects.
\subsection{Density matrix}
We will investigate here thermodynamical description treated as one-particle approximation of density operator. In real physical systems, as for example for photon gas or the system of free electrons, this is sufficient approximation to obtain satisfactory accordance with experimental data. The one-particle density operator is standardly given by occupation number operator of quantum states. For the considered case the quantum states are described by the dynamical operator basis (\ref{db}), and by this in demanded approximation the density operator has a form
\begin{equation}\label{do}
\mathsf{D}[h]={\mathsf{G}}^{\dagger}[h]{\mathsf{G}}[h].
\end{equation}
This dynamical density operator has the following matrix representation in the dynamical basis
\begin{equation}
\mathsf{D}[h]=\mathfrak{B}^{\dagger}[h]\left[\begin{array}{cc}1&0\\0&0\end{array}\right]\mathfrak{B}^{\dagger}[h]=\mathfrak{B}^{\dagger}[h]\mathbb{D}\mathfrak{B}^{\dagger}[h],
\end{equation}
and by direct application of the Bogoliubov transformation can be immediately expressed in the static initial data basis as follows
\begin{equation}
\mathsf{D}[h]=\mathfrak{B}_I^{\dagger}\left[\begin{array}{cc}|u[h]|^2&-u[h]v[h]\\-u^{\ast}[h]v^{\ast}[h]&|v[h]|^2\end{array}\right]\mathfrak{B}_I\equiv{\mathfrak{B}}_I^{\dagger}\mathbb{D}[h]{\mathfrak{B}}_I,
\end{equation}
where $\mathbb{D}[h]$ has an interpretation of the matrix representation of the density operator (\ref{do}) in the initial data operator basis, and actually describes the system of space quantum states in thermodynamical equilibrium with respect to the fundamental basis. The explicit form of the functional matrix $\mathbb{D}[h]$ is
\begin{equation}\label{den}
  \mathbb{D}[h]=\left[\begin{array}{cc}\dfrac{1}{4}\left(\sqrt[4]{\left|\dfrac{m^2}{m_I^2}\right|}+\sqrt[4]{\left|\dfrac{m_I^2}{m^2}\right|}\right)^2&\dfrac{e^{2i\theta}}{4}\left(\sqrt{\left|\dfrac{m_I^2}{m^2}\right|}-\sqrt{\left|\dfrac{m^2}{m_I^2}\right|}\right)\vspace*{10pt}\\ \dfrac{e^{-2i\theta}}{4}\left(\sqrt{\left|\dfrac{m_I^2}{m^2}\right|}-\sqrt{\left|\dfrac{m^2}{m_I^2}\right|}\right)&\dfrac{1}{4}\left(\sqrt[4]{\left|\dfrac{m^2}{m_I^2}\right|}-\sqrt[4]{\left|\dfrac{m_I^2}{m^2}\right|}\right)^2\end{array}\right],
\end{equation}
and has the natural properties
\begin{equation}
  \mathbb{D}^\dagger[h]=\mathbb{D}[h],~~\det\mathbb{D}[h]=0,
\end{equation}
where for compact notation $m=m[h]$, and $\theta=\theta[h]$. This type of reasoning is some kind of the Heisenberg picture.

It is natural to assume that the set of space quantum states is described in the Grand Canonical Ensemble \cite{ker}. The grand partition function is standardly defined as
\begin{equation}
  \Omega(z,V,T)=\tr z \exp\left(-\dfrac{U}{T}\right)=\tr \exp\left\{-\dfrac{U-\mu N}{T}\right\},
\end{equation}
where $z=\exp\dfrac{\mu N}{T}$ is called activity, $U$ is internal energy, $\mu$ is chemical potential, $N$ is averaged occupation number, $V$ is volume, and $T$ is temperature of the system. The ensemble average of quantity $A$ in the Grand Canonical Ensemble is
\begin{equation}
  \langle A\rangle =\dfrac{\tr\left(A\exp\left\{-\dfrac{U-\mu N}{T}\right\}\right)}{\tr \exp\left\{-\dfrac{U-\mu N}{T}\right\}}.
\end{equation}
Thermodynamical equation of state for the Bose system can be calculated as
\begin{equation}\label{es}
  \dfrac{PV}{T}=\ln\Omega(z,V,T),
\end{equation}
where $P$ is pressure. The famous grand partition function for the Bose statistics in the case associated with our problem is
\begin{equation}
  \Omega(z,V,T)=\dfrac{1}{1-z \exp\left(-\dfrac{U}{T}\right)},
\end{equation}
and by this the equation of state (\ref{es}) becomes
\begin{equation}\label{es1}
  \dfrac{PV}{T}=-\ln\left(1-z \exp\left(-\dfrac{U}{T}\right)\right)=\ln\dfrac{z^{-1}\exp\dfrac{U}{T}}{z^{-1}\exp\dfrac{U}{T}-1}.
\end{equation}
Moreover, the averaged occupation number can be determined
\begin{equation}\label{oc}
  N=z\dfrac{\partial}{\partial z}\ln\Omega(z,V,T)=\dfrac{z\exp\left(-\dfrac{U}{T}\right)}{1-z\exp\left(-\dfrac{U}{T}\right)}=\dfrac{1}{z^{-1}\exp\dfrac{U}{T}-1}.
\end{equation}
Entropy of the Bose gas is then determined by the following relation
\begin{equation}\label{bge}
  S=\left(\dfrac{U}{T}-\ln z\right)\dfrac{z^{-1}\exp\dfrac{U}{T}}{z^{-1}\exp\dfrac{U}{T}-1}-\ln\left(z^{-1}\exp\dfrac{U}{T}-1\right).
\end{equation}
\subsection{The Bogoliubov coefficients}
Let us consider the space quantum states system in Grand Canonical Ensemble. The basic quantity of statistical mechanics is an entropy, that for an arbitrary quantum system described is defined by the standard Gibbs--Von Neumann formula
\begin{equation}\label{entropia}
  S[h]=-\dfrac{\tr\left(\mathbb{D}[h]\ln\mathbb{D}[h]\right)}{\tr\mathbb{D}[h]},
\end{equation}
 and in considered case can be immediately computed from the density matrix (\ref{den}). Using of linear algebra methods, especially the Cayley--Hamilton characteristic polynomial and its properties, one can compute directly the logarithm of the density matrix as
 \begin{equation}
 \ln\mathbb{D}=\left[\begin{array}{cc}-\dfrac{3}{2}\dfrac{|v|^2}{|u|^2+|v|^2}+\ln\left(|u|^2+|v|^2\right)&\dfrac{5}{2}\dfrac{uv}{|u|^2+|v|^2}\\\dfrac{5}{2}\dfrac{u^\ast v^\ast}{|u|^2+|v|^2}&-\dfrac{3}{2}\dfrac{|u|^2}{|u|^2+|v|^2}+\ln\left(|u|^2+|v|^2\right)\end{array}\right],
 \end{equation}
 where $u=u[h]$ and $v=v[h]$ are the Bogoliubov coefficients given by (\ref{sup}). Taking the proper traces according to the definition (\ref{entropia}) one can directly obtain the compact relation for entropy
 \begin{equation}\label{entropy}
   S[h]=\dfrac{8|u[h]|^2|v[h]|^2}{(|u[h]|^2+|v[h]|^2)^2}-\ln\left(|u[h]|^2+|v[h]|^2\right),
 \end{equation}
that can be immediately compared with the entropy of the Bose gas (\ref{bge}), and in result leads to the following identification
\begin{eqnarray}
  |u[h]|^2+|v[h]|^2&=&z^{-1}[h]\exp\dfrac{U[h]}{T[h]}-1,\label{id1}\\
  \dfrac{8|u[h]|^2|v[h]|^2}{(|u[h]|^2+|v[h]|^2)^2}&=&\left(\dfrac{U[h]}{T[h]}-\ln z[h]\right)\dfrac{z^{-1}[h]\exp\dfrac{U[h]}{T[h]}}{z^{-1}[h]\exp\dfrac{U[h]}{T[h]}-1},\label{id2}
\end{eqnarray}
where the activity $z[h]$ is now
\begin{equation}\label{act}
  z[h]=\exp\dfrac{\mu[h]N[h]}{T[h]}.
\end{equation}
After using of the hyperbolic property of the Bogoliubov coefficients the first identification leads to the following result
\begin{equation}
z^{-1}[h]\exp\dfrac{U[h]}{T[h]}=2|u[h]|^2=2|v[h]|^2+2,
\end{equation}
and by this the second identification gives simply
\begin{equation}\label{ki}
  \dfrac{U[h]-\mu[h]N[h]}{T[h]}=\dfrac{4|v[h]|^2}{2|v[h]|^2+1}.
\end{equation}
Similarly the equation of state (\ref{es1}) for the Bose gas of space quantum states becomes
\begin{equation}\label{ik}
  \dfrac{P[h]V[h]}{T[h]}=\ln\dfrac{2|u[h]|^2}{2|u[h]|^2-1}=\ln\left(1+\dfrac{1}{2|v[h]|^2+1}\right).
\end{equation}
The formula determined averaged occupation number (\ref{oc}) expressed by the Bogoliubov coefficients becomes
\begin{equation}\label{occu}
  N[h]=\dfrac{1}{2|u[h]|^2-1}=\dfrac{1}{2|v[h]|^2+1}.
\end{equation}
The presented relations give an opportunity to determine the Helmholtz free energy
\begin{equation}\label{hel}
  F[h]=U[h]-T[h]S[h],
\end{equation}
as well as the Gibbs free energy
\begin{equation}\label{gib}
  G[h]=U[h]-T[h]S[h]+P[h]V[h],
\end{equation}
and the enthalpy of the system defined as
\begin{equation}\label{enth}
  H[h]=U[h]+P[h]V[h],
\end{equation}
iff the free energy $U[h]$ is understood as the ensemble average of the matrix representation of the Hamiltonian of the system expressed in the stable Bogoliubov vacuum
\begin{equation}\label{iedef}
  U[h]=\dfrac{\tr\mathbb{D}[h]\mathbb{H}[h]}{\tr\mathbb{D}[h]},
\end{equation}
and the thermodynamical chemical potential is simply the functional derivative of the internal energy with respect to the averaged occupation number
\begin{equation}
  \mu[h]=\dfrac{\delta U[h]}{\delta N[h]}.
\end{equation}
\subsection{Thermodynamics of the Bose gas}
In this part of the paper we will construct the space quantum states thermodynamics, that according to the conjecture presented in the first section of this text is the quantum theory of gravitation.

\begin{figure}
\centering
  \includegraphics[width=\textwidth]{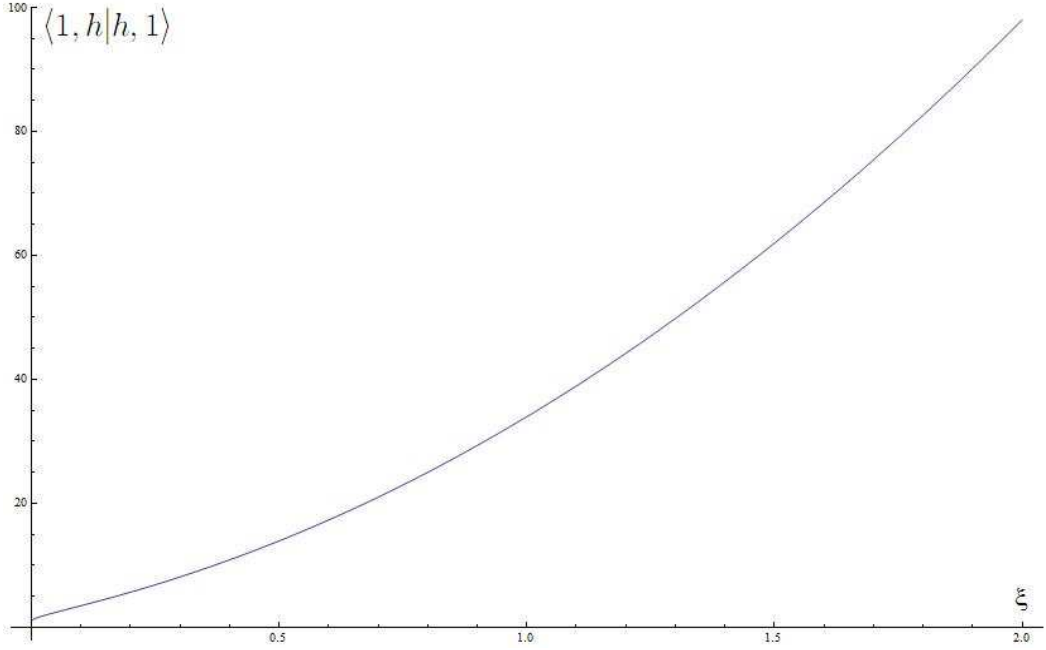}\\
  \caption{The basic one-point correlation function for stable configuration of space quantum states. For the classical limit, \emph{i.e.} for huge number of vacuum quantum states, the one-point correlations arises infinitely.}\label{fkor}
\end{figure}

Let us start from derivation of the thermodynamical quantities for the Bose gas of space quantum states that give crucial information about this many-body statistical system. Firstly, we will consider the internal energy of the gas. In order to derivation this characteristics, let us consider the matrix representation $\mathbb{H}$ of the quantum field theory Hamiltonian (\ref{qfth}) of the space quantum states with respect to the initial data fundamental operator basis $\mathfrak{B}_I$, that is
\begin{equation}
  \mathbb{H}[h]=\left[\begin{array}{cc}\dfrac{m[h]}{2}\left(|v[h]|^2+|u[h]|^2\right)&-m[h]u[h]v[h]\\-m[h]u^\ast[h]v^\ast[h]&\dfrac{m[h]}{2}\left(|v[h]|^2+|u[h]|^2\right)\end{array}\right].
\end{equation}
As it can be checked directly, this Hamiltonian matrix for fixed space metrics has the discrete spectrum that consists two different type eigenvalues
\begin{equation}\label{eig}
  \mathrm{Spec}\mathbb{H}=\left\{\dfrac{m[h]}{2}\left(|v[h]|+\sqrt{1+|v[h]|^2}\right)^2,\dfrac{m[h]}{2}\left(|v[h]|-\sqrt{1+|v[h]|^2}\right)^2\right\}.
\end{equation}
By using of the definition (\ref{iedef}) and some elementary algebraic computations one can obtain directly the internal energy of the Bose gas, that is equal to
\begin{equation}\label{int1}
  U[h]=m[h]\left(|v[h]|^2+\dfrac{1}{2}+\dfrac{|v[h]|^2\left(1+|v[h]|^2\right)}{|v[h]|^2+\dfrac{1}{2}}\right).
\end{equation}
Let us concentrate our attention on the occupation number of quantum states for the considered Bose gas of space quantum states. The number of space quantum states generated from the stable Bogoliubov vacuum related to initial data fundamental operator basis can be derived by standard method, as the vacuum expectation value of the one--particle density operator (\ref{do}), namely by the following way
\begin{equation}
 \xi=\dfrac{{_I}\langle0\left|\mathsf{D}[h]\right|0\rangle_I}{{_I}\langle0|0\rangle_I}=\dfrac{{_I}\langle0\left|{\mathsf{G}}^{\dagger}[h]{\mathsf{G}}[h]\right|0\rangle_I}{{_I}\langle0|0\rangle_I}.
\end{equation}
After direct application of the bosonic Bogoliubov transformation and by using of the canonical commutation relations related to the fundamental initial data operator basis in the Fock space, one can simply derive the number of vacuum quantum states as follows
\begin{eqnarray}\label{nof}
  &&\xi=\dfrac{{_I}\left\langle0\left|\left(u[h]\mathsf{G}_I^\dagger-v[h]\mathsf{G}_I\right)\left(-v^\ast[h]\mathsf{G}_I^\dagger+u^\ast[h]\mathsf{G}_I\right)\right|0\right\rangle_I}{{_I}\langle0|0\rangle_I}\nonumber=\\
  &&=\dfrac{{_I}\left\langle0\left||v[h]|^2\mathsf{G}_I\mathsf{G}_I^\dagger+|u[h]|^2\mathsf{G}_I^\dagger\mathsf{G}_I-v^\ast[h]u[h]\mathsf{G}_I^\dagger\mathsf{G}_I^\dagger-v[h]u^\ast[h]\mathsf{G}_I\mathsf{G}_I\right|0\right\rangle_I}{{_I}\langle0|0\rangle_I}\nonumber=\\
  &&=|v[h]|^2\dfrac{{_I}\left\langle0\left|\mathsf{G}_I\mathsf{G}_I^\dagger\right|0\right\rangle_I}{{_I}\langle0|0\rangle_I}=|v[h]|^2.
\end{eqnarray}
From the other point of view one can calculate the number of all possible states that can be occupied by the ensemble. This quantity can be determined by grand canonical ensemble average of the matrix representation of one--particle density operator according to the definition
\begin{equation}
  \langle N\rangle[h]=\dfrac{\tr\left(\mathbb{D}[h]\mathbb{N}[h]\right)}{\tr\mathbb{D}[h]}.
\end{equation}
However, in the considered case we have the identification $\mathbb{N}[h]\equiv\mathbb{D}[h]$, and by this reason the number can be computed immediately with the following result
\begin{equation}
  \langle N\rangle[h]=\dfrac{\tr\left(\mathbb{D}^2[h]\right)}{\tr\mathbb{D}[h]}=\dfrac{\tr\left((\tr\mathbb{D}[h])\mathbb{D}[h]\right)}{\tr\mathbb{D}[h]}=\tr\mathbb{D}[h],
\end{equation}
that after application of the matrix representation (\ref{den}) and the relation (\ref{nof}) leads finally to
\begin{equation}\label{nn}
  \langle N\rangle=2\xi+1.
\end{equation}
By this the grand canonical ensemble average of an occupation number of the Bose gas of space quantum states determined firstly by the relation $(\ref{occu})$ really equals
\begin{equation}\label{oc1}
  N=\dfrac{1}{2\xi+1},
\end{equation}
and gives an information that statistically the volume of the Bose gas of space quantum states is occupied by one space quantum state. The relation between the number of states generated from the stable Bogoliubov vacuum $\xi$ and the mass $m[h]$ arises directly from the formula (\ref{sup2}) as
\begin{equation}\label{maas}
  m_{\pm}=m_I\left(\sqrt{\xi}\pm\sqrt{\xi+1}\right)^2.
\end{equation}
By this reason the spectrum of the Hamiltonian eigenvalues (\ref{eig}) actually is determined by
\begin{equation}\label{eig}
  \mathrm{Spec}\mathbb{H}=\left\{\dfrac{m_I}{2}\left(\sqrt{\xi}+\sqrt{\xi+1}\right)^4,\dfrac{m_I}{2}\left(\sqrt{\xi}-\sqrt{\xi+1}\right)^4\right\}.
\end{equation}
This result can be interpreted as follows -- the Bose gas of space quantum states consists two physically independent phases, associated with the sign $+$ and $-$ respectively. However, these two possible phases have no the same physical status. For demystify of this fact let us consider the basic quantity (\ref{cor2}) of the quantum field theory formulated in previous parts of this paper -- the correlation function, that carries an information about one-point bosonic field configuration and is the key quantity by this fact
\begin{equation}\label{opc}
  \langle1h|h1\rangle_{\pm}=\left(\dfrac{m_I}{m_{\pm}}\right)^2=\dfrac{1}{\left(\sqrt{\xi}\pm\sqrt{\xi+1}\right)^4}.
\end{equation}
The character of changeability of this one-point correlator strongly depends on the choice of the sign in the denominator, and has completely different physical meaning for the case of the sign $+$ and for the case of the sign $-$. Namely, in the case the positive sign this correlator goes to zero for huge values of particles generated from the initial data vacuum, but for the case of negative sign the one-point correlations become asymptotically infinite for huge number of vacuum quantum states
\begin{equation}
  \lim_{\xi\rightarrow\infty}\langle1h|h1\rangle_{\pm}=\left\{\begin{array}{cc}0&~,~~\mathrm{for}~~$+$\\
  \infty&,~~\mathrm{for}~~$--$\end{array}\right.
\end{equation}
The physical meaning of this situation can be explained in the following way. In the case of the positive sign, the one-point correlations in the limit of huge number of quantum states generated from the stable Bogoliubov vacuum asymptotically vanish, that physically means we have to deal with unstable situation in the classical limit, and by this reason the classical object associated with the positive sign in the one-point correlator (\ref{opc}) is the unstable object. However, in the second case, that is for the negative sign, the one-point correlations asymptotically arise to infinity with arise to infinity of the number of vacuum quantum states, and by this reason in this case the one-point correlator (\ref{opc}) describes stable configuration of space quantum states in the classical limit, it is stable physical object, see Figure (\ref{fkor}). In this manner, at the present text we will discuss only the case of the negative sign.
\pagebreak
\subsection{Classically stable phase. Cold Big Bang.}
\begin{figure}
\centering
  \includegraphics[width=\textwidth]{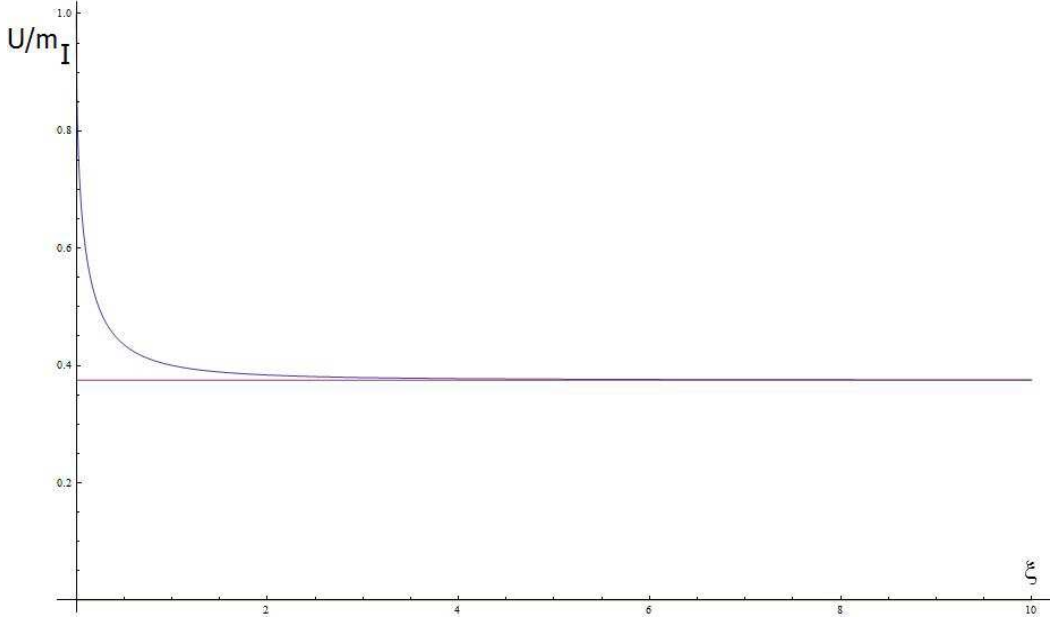}\\
  \caption{Internal energy for the Bose gas of space quantum states primordially is given by initial data, and asymptotically goes to constant value determined also by initial data.}\label{f1}
\end{figure}

\begin{figure}
  \includegraphics[width=\textwidth]{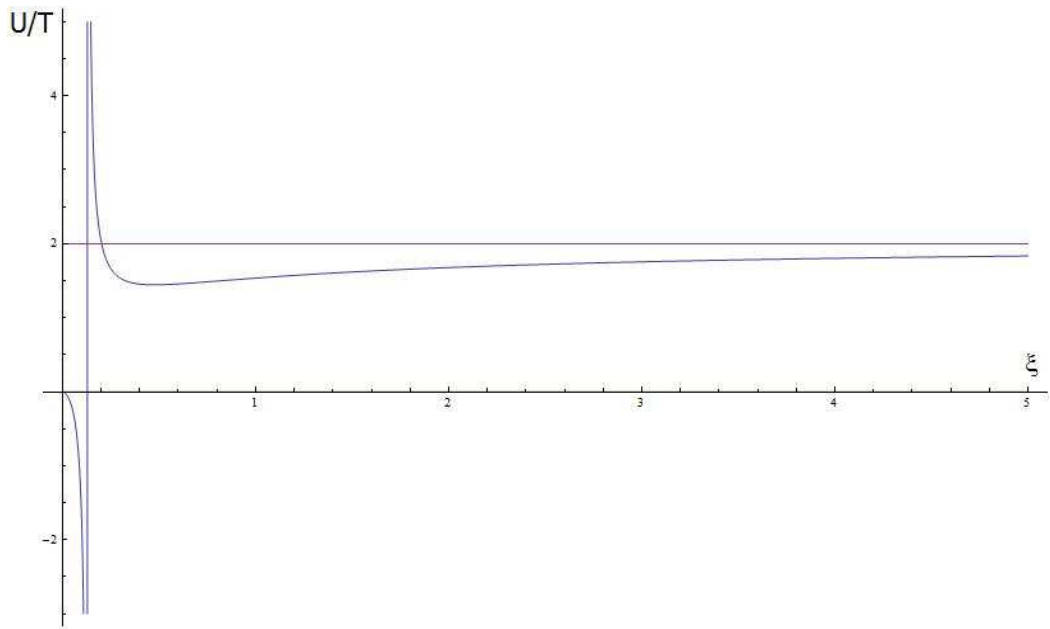}\\
  \caption{Relation between quotient of internal energy and temperature (the blue line), and number of space quantum states generated from the initial data Bogoliubov vacuum. For the limit of huge value of this quantum number (the red line), \emph{i.e.} for the classical equilibrium state of the system of space quantum states described by a Riemannian manifold given by a solution of the Einstein--Hilbert field equations of General Relativity, according to the law of equipartition this quotient asymptotically is related to 4 degrees of freedom, which have an interpretation of four spacetime coordinates.}\label{ut}
\end{figure}

The internal energy of the Bose gas of space quantum states (\ref{int1}) for the stable fields configuration reads
\begin{equation}\label{int2}
   U=m_I\dfrac{3\xi^2+3\xi+1}{2\xi+1}\left(\sqrt{\xi}-\sqrt{\xi+1}\right)^2,
\end{equation}
and is monotonic function of the argument $\xi$ (see Figure (\ref{f1})) that in classical limit of the huge argument values goes asymptotically to the constant value that is
\begin{equation}
  \lim_{\xi\rightarrow\infty}U=\dfrac{3}{8}m_I.
\end{equation}
\begin{figure}
  \includegraphics[width=\textwidth]{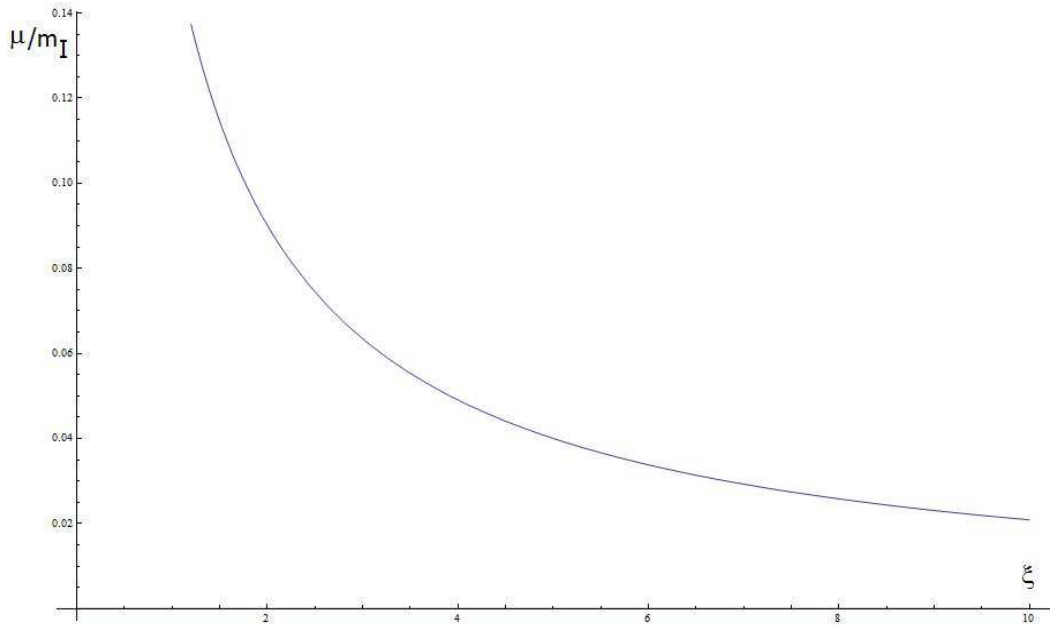}\\
  \caption{Chemical potential for the Bose gas of space quantum states asymptotically describes open system, but primordially is associated with a point object ( The Big Bang point).}\label{f2}
\end{figure}
\begin{figure}
  \includegraphics[width=\textwidth]{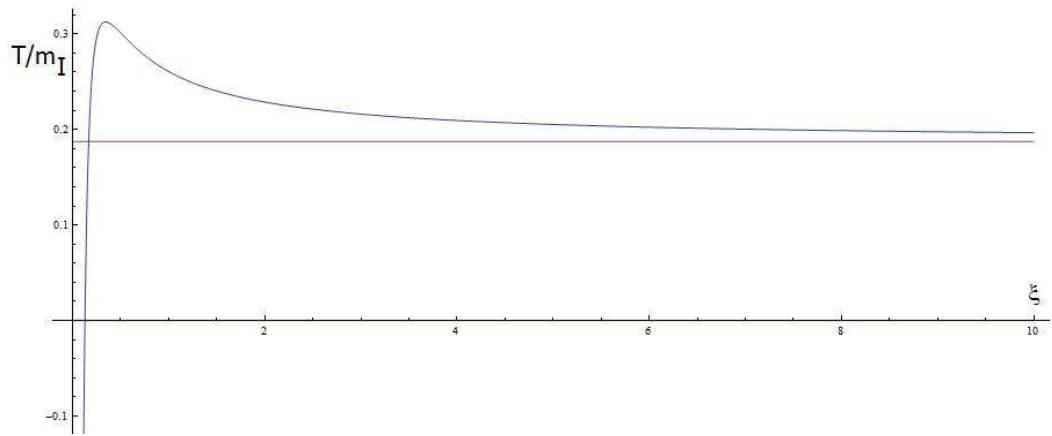}\\
  \caption{Temperature for the Bose gas of space quantum states asymptotically goes to constant value determined by initial data, but primordially in the point of Big Bang is characterized by minus infinite value of temperature (Cold Big Bang).}\label{f3}
  \end{figure}
One can characterize some statistical properties of the Bose gas of space quantum states by derivation of the chemical potential for this system. This quantity can be calculated by direct using of the standard thermodynamical relation
\begin{equation}\label{che}
  \mu=\dfrac{\delta U}{\delta N},
\end{equation}
that by using of the fact that the internal energy $U$ as well as the averaged occupation number $N$ are functions of the number of vacuum quantum states $\xi$ leads to the definition
\begin{equation}\label{che}
  \mu=\dfrac{\delta U}{\delta\xi}\dfrac{\delta \xi}{\delta N}.
\end{equation}
By using of the relations (\ref{int2}) and (\ref{oc1}) one can compute some elementary derivatives that are need for derivation of the chemical potential
\begin{eqnarray}
  \dfrac{\delta U}{\delta\xi}&=&m_I\left[\dfrac{6\xi^2+6\xi+1}{2\xi+1}-\dfrac{3\xi^2+3\xi+1}{\sqrt{\xi(\xi+1)}}\right]\dfrac{\left(\sqrt{\xi}-\sqrt{\xi+1}\right)^2}{2\xi+1},\\
  \dfrac{\delta \xi}{\delta N}&=&\left(\dfrac{\delta N}{\delta \xi}\right)^{-1}=-\dfrac{1}{2}(2\xi+1)^2,
\end{eqnarray}
so that actually the chemical potential (\ref{che}) depends from number of vacuum quantum states by the following formula
\begin{eqnarray}\label{pch}
  \mu=-m_I\left[3\xi^2+3\xi+\dfrac{1}{2}-\dfrac{3\xi^2+3\xi+1}{2\sqrt{\xi(\xi+1)}}\left(2\xi+1\right)\right]\left(\sqrt{\xi}-\sqrt{\xi+1}\right)^2.
\end{eqnarray}
The chemical potential (\ref{pch}) is also monotonic function of the argument $\xi$ and asymptotically decreases to zero for huge number of vacuum particles. This fact physically means that we actually consider the system with nonconserved number of quantum states (see Figure (\ref{f2})). Obviously, it is not new fact for our considerations, we have considered this type system in this paper from the beginning. Now temperature of the Bose gas of space quantum states can be determined by direct application of the relation (\ref{ki}) as follows
\begin{eqnarray}
T=\dfrac{2\xi+1}{4\xi}\left(U-\mu N\right),
\end{eqnarray}
that after application of the relations (\ref{int2}), (\ref{oc1}), and (\ref{pch}) leads to the relation between  temperature and  number of the space quantum states produced from initial data vacuum
\begin{equation}
  T=m_I\left[4\xi^2+4\xi+1-\dfrac{3\xi^2+3\xi+1}{\sqrt{\xi(\xi+1)}}(2\xi+1)\right]\dfrac{3\left(\sqrt{\xi}-\sqrt{\xi+1}\right)^2}{8\xi}.
\end{equation}
This temperature globally is not monotonic function, but has stable value in the classical limit (see Figure (\ref{f3}))
\begin{equation}
  \lim_{\xi\rightarrow\infty}T=\dfrac{3}{16}m_I.
\end{equation}
One can see now that the following relation between internal energy and temperature of space quantum states holds
\begin{equation}\label{utr}
  \dfrac{U}{T}=\dfrac{\dfrac{8}{3}\dfrac{\xi}{2\xi+1}}{\dfrac{4\xi^2+4\xi+1}{3\xi^2+3\xi+1}-\dfrac{2\xi+1}{3\sqrt{\xi(\xi+1)}}},
\end{equation}
and in the limit of huge number of vacuum quantum states one can suppose that the principle of energy equipartition should be fulfilled -- in a sense of the classical thermal equilibrium, the energy is shared equally among on all degrees of freedom $f$ of the system
\begin{equation}\label{epl}
\xi\rightarrow\infty \Longrightarrow U=\dfrac{f}{2}T.
\end{equation}
One can calculate immediately the classical limit of the relation (\ref{utr}). The result exactly accords with the equipartition law (\ref{epl}), for this case the number of degrees of freedom equals
\begin{equation}
  f=4.
\end{equation}
For huge number of vacuum quantum states we have to deal with classical thermal equilibrium state of the system of space quantum states, that is a Riemannian manifold given by a solution of the Einstein--Hilbert field equations of General Relativity. Simultaneously out of the presented way looks into view the following fact: \emph{classical thermal equilibrium state of the system of space quantum states is associated with an object described by 4 thermodynamical degrees of freedom} (see Figure (\ref{ut})). These degrees of freedom have the natural interpretation -- they can be identified with four spacetime coordinates - one time and three space coordinates.

\begin{figure}
  \includegraphics[width=\textwidth]{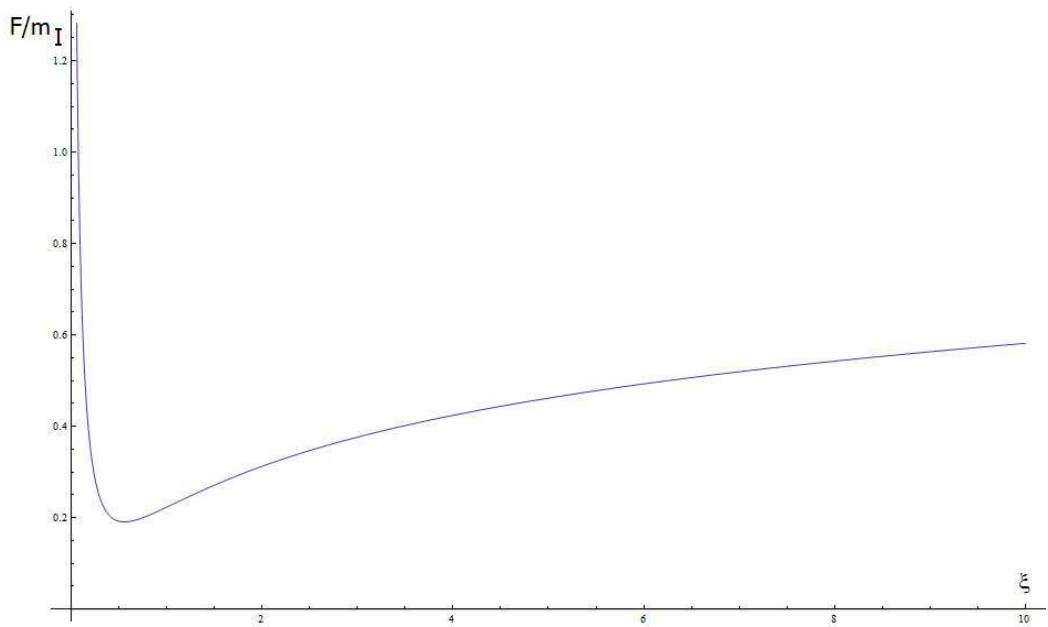}\\
  \caption{Helmholtz free energy for the Bose gas of space quantum states asymptotically arises to infinity.}\label{f5}
  \end{figure}

\begin{figure}
  \includegraphics[width=\textwidth]{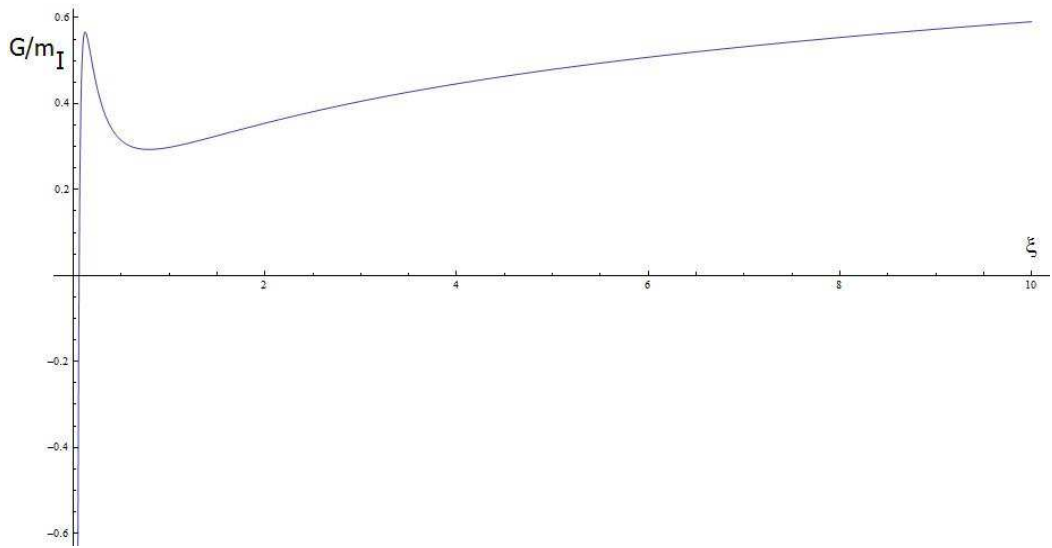}\\
  \caption{Gibbs free energy for the Bose gas of space quantum states asymptotically arises to infinity.}\label{f6}
\end{figure}

\begin{figure}
  \includegraphics[width=\textwidth]{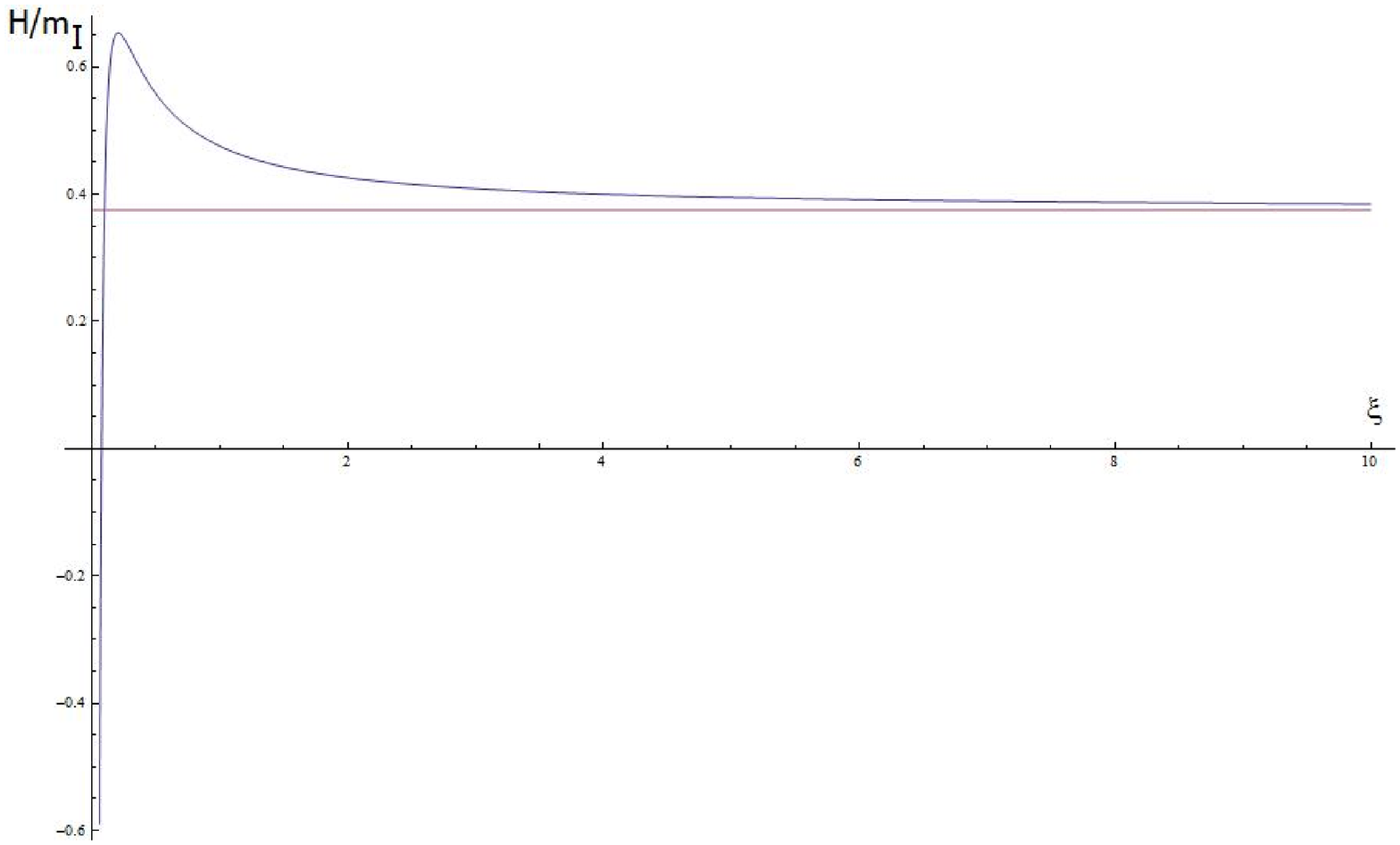}\\
  \caption{Enthalpy for the Bose gas of space quantum states asymptotically goes to constant value determined by initial data.}\label{f7}
\end{figure}

By direct using the equation of state (\ref{ik}) one can determinate the product of pressure and volume as
\begin{eqnarray}\label{pre}
  PV&=&m_I\left[\dfrac{4\xi^2+4\xi+1}{2\xi+1}-\dfrac{3\xi^2+3\xi+1}{\sqrt{\xi(\xi+1)}}\right]\times\nonumber\\
  &\times&\dfrac{3(2\xi+1)}{8\xi}\left(\sqrt{\xi}-\sqrt{\xi+1}\right)^2\ln\left(\dfrac{2\xi+2}{2\xi+1}\right),
\end{eqnarray}
and similarly the product of temperature and entropy as
\begin{eqnarray}\label{ts}
  TS&=&3m_I\left[\dfrac{4\xi^2+4\xi+1}{2\xi+1}-\dfrac{3\xi^2+3\xi+1}{\sqrt{\xi(\xi+1)}}\right]\times\nonumber\\&\times&\left[\dfrac{\xi+1}{2\xi+1}-\dfrac{2\xi+1}{8\xi}\ln(2\xi+1)\right]\left(\sqrt{\xi}-\sqrt{\xi+1}\right)^2.
\end{eqnarray}
The product of pressure and volume (\ref{pre}) goes to zero for huge number of vacuum space quantum states, what physically means that in this limit the pressure goes to zero for arbitrary big volume. The product of entropy and temperature (\ref{ts}) goes to minus infinity in this limit. Now the Helmholtz free energy $F$ given by general relation (\ref{hel}) can be determined directly as follows
\begin{eqnarray}
  F&=&m_I\Biggr\{1+3\left[\dfrac{2\xi+1}{8\xi}\ln(2\xi+1)-\dfrac{\xi+1}{2\xi+1}\right]\dfrac{3\xi^2+3\xi+1}{2\xi+1}\times\nonumber\\
  &\times&\left[\dfrac{4\xi^2+4\xi+1}{3\xi^2+3\xi+1}-\dfrac{2\xi+1}{\sqrt{\xi(\xi+1)}}\right]\Biggr\}\left(\sqrt{\xi}-\sqrt{\xi+1}\right)^2,
\end{eqnarray}
similarly the Gibbs free energy $G$ determined standardly by the relation (\ref{gib}) now is equal to
\begin{eqnarray}
  G&=&m_I\Biggr\{1+3\left[\dfrac{2\xi+1}{8\xi}\ln(2\xi+2)-\dfrac{\xi+1}{2\xi+1}\right]\dfrac{3\xi^2+3\xi+1}{2\xi+1}\times\nonumber\\
  &\times&\left[\dfrac{4\xi^2+4\xi+1}{3\xi^2+3\xi+1}-\dfrac{2\xi+1}{\sqrt{\xi(\xi+1)}}\right]\Biggr\}\left(\sqrt{\xi}-\sqrt{\xi+1}\right)^2,
\end{eqnarray}
and the enthalpy $H$ defined by the formula (\ref{enth}) now reads
\begin{eqnarray}
  H&=&m_I\Biggr\{1+\dfrac{3(2\xi+1)}{8\xi}\left[\dfrac{4\xi^2+4\xi+1}{3\xi^2+3\xi+1}-\dfrac{2\xi+1}{\sqrt{\xi(\xi+1)}}\right]\times\nonumber\\
  &\times&\ln\left(\dfrac{2\xi+2}{2\xi+1}\right)\Biggr\}\dfrac{3\xi^2+3\xi+1}{2\xi+1}\left(\sqrt{\xi}-\sqrt{\xi+1}\right)^2.
\end{eqnarray}
These thermodynamical potentials have the following asymptotical values for huge number of vacuum quantum states (see Figures (\ref{f5}), (\ref{f6}), and (\ref{f7}))
\begin{eqnarray}
\lim_{\xi\rightarrow\infty}F&=&\infty,\\
\lim_{\xi\rightarrow\infty}G&=&\infty,\\
\lim_{\xi\rightarrow\infty}H&=&\dfrac{3}{8}m_I.
\end{eqnarray}
Above thermodynamical characteristics determine complete physical information about the Bose gas of space quantum states related to the initial data stable Bogoliubov vacuum state. The variable $\xi$, that really is a number of space quantum states generated from the stable vacuum and simultaneously the square of one of the Bogoliubov coefficients, can be treated as the fundamental quantity directly related with the basic one-point correlator $\langle1h|h1\rangle$ by the following way
\begin{equation}
  \xi=\dfrac{1}{4}\left(\dfrac{1}{\sqrt{\left|\langle1h|h1\rangle\right|}}+\sqrt{\left|\langle1h|h1\rangle\right|}\right)-\dfrac{1}{2},
\end{equation}
and allows to study the presented relations between thermodynamics of space quantum states and classical equilibrium states determined only by the one-point correlator.
\subsection{Entropy}
\begin{figure}
\centering
  \includegraphics[width=\textwidth]{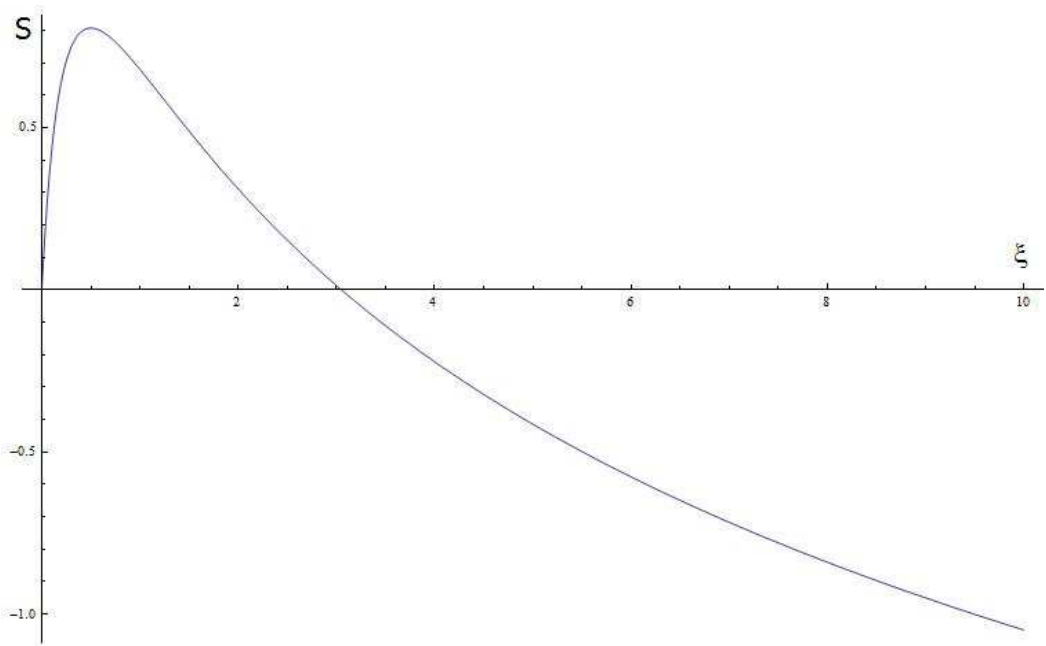}\\
  \caption{Entropy of the Bose gas of space quantum states as a function of the parameter $\xi$ has nontrivial maximum, that is identified with the Bose condensation in the system.}\label{fig:ent}
\end{figure}

Let us consider the entropy of the Bose gas of space quantum states (\ref{entropy}), graphically presented on the Figure (\ref{fig:ent}).
The relation (\ref{entropy}) actually establishes the nontrivial connection between disorder in the Bose gas of space quantum states with respect to the initial data fundamental operator basis, and the number of space quantum states generated from the stable initial data vacuum state. Let us take into our considerations the set of initial data space quantum states. According to the relation for the mass (\ref{maas}) this group of space quantum states is described by the initial data mass
\begin{equation}
  m=m_I\Longleftrightarrow\xi=0,
\end{equation}
that really is the initial tachyon mass. It can be seen directly that the entropy (\ref{entropy}) for these quantum states vanishes
\begin{equation}
S_I=0.
\end{equation}
The complete thermodynamical characterizations of the group of initial data space quantum states can be computed by taking the limit
\begin{eqnarray}
  \lim_{\xi\rightarrow0}N&=&1,\\
  \lim_{\xi\rightarrow0}U&=&m_I,\\
  \lim_{\xi\rightarrow0}\mu&=&\infty,\\
  \lim_{\xi\rightarrow0}T&=&-\infty,\\
  \lim_{\xi\rightarrow0}\dfrac{PV}{T}&=&\ln2,\\
  \lim_{\xi\rightarrow0}F&=&\infty,\\
  \lim_{\xi\rightarrow0}G&=&-\infty,\\
  \lim_{\xi\rightarrow0}H&=&-\infty.
\end{eqnarray}
We see that initially the Bose gas of space quantum states has finite internal energy and unit averaged occupation number, but all other characterizations are $\pm$ infinite. The chemical potential is also infinite, that physically means that the system is no open and is compact point object with huge negative temperature. In this manner the initial data point can be interpreted as the Big Bang point, where objects that bangs are space quantum states spontaneously generated from the stable quantum vacuum. The temperature in the Big Bang limit is negative infinite; this phenomena can be called the Cold Big Bang.

On the Figure (\ref{fig:ent}) we see that the next interesting group of states are the space quantum states associated with the maximal value of the entropy. Let us consider now the maximally entropy point of the system of space quantum states. This especial point is determined by the number of quantum states generated from the initial data vacuum given by
\begin{equation}
  \xi=\dfrac{1}{2},
\end{equation}
and by maximal value of entropy and chemical equilibrium character, this point has the natural interpretation of the point of the condensation in the Bose gas of space quantum states. The averaged occupation number for the condensate state equals
 \begin{equation}
   N_{cond}=\dfrac{1}{2}=\xi,
 \end{equation}
and the mass of the condensate is
 \begin{equation}
   m_{cond}\approx0.26795m_I.
 \end{equation}
The Bose condensate point has the entropy
\begin{equation}
 S_{cond}=\dfrac{3}{2}-\ln2,
\end{equation}
and by values of thermodynamical characteristics are as follows
\begin{eqnarray}
  U_{cond}&\approx&0.43542m_I,\\
  \mu_{cond}&\approx&0.26869m_I,\\
  T_{cond}&\approx&0.30107m_I,\\
  F_{cond}&\approx&0.19250m_I,\\
  G_{cond}&\approx&0.31457m_I,\\
  H_{cond}&\approx&0.55749m_I,
\end{eqnarray}
with the following equation of state
\begin{equation}
    \left(\dfrac{PV}{T}\right)_{cond}=\ln\dfrac{3}{2}.
\end{equation}

The group of space quantum states with maximal value of entropy are formally associated with chemical equilibrium of the Bose system of space quantum states. The fact that the maximal value of entropy is not localized in the initial data point $\xi=0$ is the typical characteristic property of systems with the Bose condensate presence. However, as it was presented the group of initial data space quantum states play the crucial role in context of the Einstein--Hilbert General Relativity Riemannian manifold. These states have the fundamental status, they have the primordial states meaning. From the Figure (\ref{fig:ent}) we see that in the region between the Big Bang and the Bose condensation of space quantum states, \emph{i.e.} $0\leq\xi\leq\dfrac{1}{2}$, we observe entropy arising, and from the Bose condensation point up to classical equilibrium state, \emph{i.e.} in the region $\xi\geq\dfrac{1}{2}$, entropy decreases to minus infinity, and system goes to thermodynamical disorder. It is interesting that in this region exists the point when entropy again vanishes, \emph{i.e} the point where the system of space quantum states has the initial value of entropy, but other characteristics are not the same as initial ones.
\newpage
\section{Summary}
In this paper we have presented is details the new realization of the old problem, that is formulation of quantum gravity by effective thermodynamics of quantum states. This realization was based on the fundamental fact -- the Wheeler--DeWitt theory following from $3+1$ decomposition of a Lorentzian manifold metric field of General Relativity actually is not nonrelativistic Schr\"odinger quantum mechanics for wave function of Universe, but is the Global One--Dimensional Klein--Gordon--Fock equation of classical field theory of the Bose field associated with embedded 3-dimensional space. Moreover, we have proved directly that the Wheeler-DeWitt equation with presence superspace metrics can be represented in the form without explicitly using of the superspace metrics, that is the idea of Global One-Dimensionality, where the dimension is 3-volume form of a space. This simplified equation that further exists in the configurational space of General Relativity, was treated in this paper as the equation of classical field theory of the Bose field associated with spatial geometry of the Einstein--Hilbert pseudo--Riemannian manifold. The tachyon state of the classical field theory and more important from physical point of view the Dark Energy and the cosmological constant problems were described and discussed. Some limits for value of cosmological constant was derived. This little interpretational and cosmetic changes in the form of the Wheeler--DeWitt theory completely changed essence of the geometrodynamics that in the presented form is a classical field theory of some relativistic system. Quantization of this classical field theory is natural by application of the language of the Fock space functional annihilation and creation operators, and as it was presented in this paper gives beautiful and elegant results on physical nature of quantum gravitation. The quantum field theory was used as the main link between the Einstein--Hilbert General Relativity and thermodynamical description of a Lorentzian manifold as an effect of the many-body quantum field theory of the Bose gas of space quantum states. It is the general field theory according to depictions of Dr. Albert Einstein.

The quantum theory of gravitation presented in this paper essentially describes the classical spacetime given by a solution of General Relativity field equations as an effect of asymptotical equilibrium of the generalized thermodynamics of the Bose gas of many quantum states of three-dimensional space that classically evolves in 1-dimensional time. These quantum states of $3+1$ splitted pseudo--Riemannian spacetime was called in this paper by name of space quantum states, and it was shown here that these quantum states can be considered in terms of gravity quanta. These are gravitons, in the sense of quantum field theory formulated in the Fock space the stable Bogoliubov vacuum state. As it was seen, the quantization of classical field theory, given by one-dimensional Klein--Gordon--Fock equation, can be constructed correctly only by using of the Fock space operator basis that is the Heisenberg type, \emph{i.e.} has static character. This fact caused using of the bosonic Bogoliubov transformation, and leads to the fundamental operator basis associated with initial data. However, still we have to deal with open system, where number of quantum states is not conserved. The description related to the stable Bogoliubov vacuum state gave an opportunity to understand the quantum field theory in terms of the Bose gas of space quantum states and construct proper statistical mechanics of this system. As it was shown explicitly, this amazing Bose gas has some state of chemical equilibrium that is related to non-zero number of quantum states generated from the initial data vacuum, and has an interpretation of the Bose condensation in the Bose gas of space quantum states with respect to the initial data vacuum state. Really, as it was mentioned the Bose gas of space quantum consists two physical phases -- one phase is characterized by some kind of "condensation" of one-point correlations in the case of arising of number of quantum states generated from the fundamental initial data vacuum, the second phase has vanishing correlations of the Bose gas in this classical limit. Physically this fact means that classical solution given by a Lorentizan manifold of General Relativity is described by the stable classical equilibrium in the first case, and by completely unstable state in the case of the second phase. By this physical reasoning we have chosen to further consideration the phase with condensing one-point correlations. This solution is thermodynamically stable in the classical limit, \emph{i.e.} in the limit of huge value of number of vacuum quantum states gives a classical object that can be identified with the pseudo--Riemannian spacetime. Furthermore, as it was computed by using of the equipartition law, this solution gives a number of thermodynamical degrees of freedom which accords to number of classical spacetime coordinates - it is equal to four. By this reason, from the point of view of the thermodynamics of the Bose gas of space quantum states, spacetime coordinates have a status of thermodynamical degrees of freedom in the presented approach. In this classical limit the considered phase has asymptotically constant values of temperature and internal energy, and chemical potential vanishes classically. In this manner the stable solution describes the classical open quantum system in constant temperature. In the case of the second solution, by vanishing of the one-point correlations in classical limit, also the temperature of the space quantum states system arises to infinity for huge number of vacuum quantum states. For this solution exists Hot Big Bang of the initial data quantum states from the Bogoliubov vacuum opportunity. As it was presented in details, the analysis of the stable phase of the Bose gas of space quantum states leads to the Cold Big Bang phenomena, and to the interpretation of the initial state of the Bose gas as the primordial compact point object with negative infinite temperature.

Finally, it was shown that the condensed state of the Bose gas of space quantum states presented for nontrivial value of number of vacuum quantum states, has a natural physical interpretation of chemical equilibrium state. Actually, by application of an analogy famous from condensed matter physics, the Bose condensate has a nature of the quantum object with macroscopic dimensions, and this should be observed in experiments as in the case of the Bose--Einstein condensation of photons. The condensed state of the Bose gas of space quantum states can be responsible for Dark Matter effects.

\section*{Acknowledgements}
Special thanks are directed to Dr. B. G. Sidharth and Prof. F. Honsell for full hospitality 3 - 25 June 2008 at Dipartimento di Matematica e Informatica of Universit$\grave{\mathrm{a}}$ degli Studi di Udine and discussions during the stay.

The author benefitted many valuable discussions from Prof. G. 't Hooft and Dr. B. G. Sidharth, and is grateful to Profs. A. B. Arbuzov, I. Ya. Aref'eva, B. M. Barbashov, K. A. Bronnikov, I. L. Buchbinder, V. N. Pervushin, and V. B. Priezzhev for the critical remarks and motivation.

\end{document}